\documentclass[fleqn,usenatbib]{mnras}

\usepackage{newtxtext,newtxmath}

\usepackage[T1]{fontenc}
\usepackage{booktabs}
\usepackage{wrapfig}
\usepackage{subfig}

\DeclareRobustCommand{\VAN}[3]{#2}
\let\VANthebibliography\thebibliography
\def\thebibliography{\DeclareRobustCommand{\VAN}[3]{##3}\VANthebibliography}

\usepackage{graphicx}	
\usepackage{svg}

\newcommand{\arepo}{{\sc Arepo}}
\newcommand{\smuggle}{\textit{SMUGGLE}}

\title[Stellar feedback and gas accretion]{Galactic coronae in Milky Way-like galaxies: the role of stellar feedback in gas accretion}

\author[F. Barbani et al.]{Filippo Barbani$^{1,2}$\thanks{E-mail: filippo.barbani2@unibo.it},
Raffaele Pascale$^{2}$,
Federico Marinacci$^{1,2}$,
Laura V. Sales$^{3}$,\newauthor
Mark Vogelsberger$^{4,5}$,
Paul Torrey$^{6}$,
Hui Li$^{7,8}$\\
$^{1}$Department of Physics and Astronomy, University of Bologna, Via P. Gobetti 93/2, I-40129 Bologna, Italy\\
$^{2}$INAF, Astrophysics and Space Science Observatory Bologna, Via P. Gobetti 93/3, I-40129 Bologna, Italy\\
$^{3}$Department of Physics and Astronomy, University of California, Riverside, CA 92521, USA\\
$^{4}$Department of Physics and Kavli Institute for Astrophysics and Space Research, Massachusetts Institute of Technology, Cambridge, MA 02139, USA\\
$^{5}$The NSF AI Institute for Artificial Intelligence and Fundamental Interactions, Massachusetts Institute of Technology, Cambridge, MA 02139, USA\\
$^{6}$Department of Astronomy, University of Florida, 211 Bryant Space Science Center, Gainesville, FL 32611, USA\\
$^{7}$Department of Astronomy, Columbia University, New York, NY 10027, USA\\
$^{8}$Department of Astronomy, Tsinghua University, Beijing 100084, China\\
}

\date{Accepted XXX. Received YYY; in original form ZZZ}

\pubyear{2023}

\begin{document}
\label{firstpage}
\pagerange{\pageref{firstpage}--\pageref{lastpage}}
\maketitle

\begin{abstract}
Star-forming galaxies like the Milky Way are surrounded by a hot gaseous halo at the virial temperature -- the so-called galactic corona -- that plays a fundamental role in their evolution. The interaction between the disc and the corona has been shown to have a direct impact on accretion of coronal gas onto the disc with major implications for galaxy evolution. 
In this work, we study the gas circulation between the disc and the corona of star-forming galaxies like the Milky Way. We use high-resolution hydrodynamical $N$-body simulations of a Milky Way-like galaxy with the inclusion of an observationally-motivated galactic corona. In doing so, we use \smuggle, an explicit interstellar medium (ISM) and stellar feedback model coupled with the moving-mesh code \arepo. We find that the reservoir of gas in the galactic corona is sustaining star formation: the gas accreted from the corona is the primary fuel for the formation of new stars, helping in maintaining a nearly constant level of cold gas mass in the galactic disc. Stellar feedback generates a gas circulation between the disc and the corona (the so-called galactic fountain) by ejecting different gas phases that are eventually re-accreted onto the disc. The accretion of coronal gas is promoted by its mixing with the galactic fountains at the disc-corona interface, causing the formation of intermediate temperature gas that enhance the cooling of the hot corona. We find that this process acts as a positive feedback mechanism, increasing the accretion rate of coronal gas onto the galaxy.

\end{abstract}




\section{Introduction}\label{introduction}
Galaxies like the Milky Way have formed stars with an almost constant star formation rate for about their entire life \citep{twarog1980,cignoni2006, bernard2018, alzate2021}. However, the mass of gas located in their discs is not sufficient to sustain their current star formation rate (i.e. $\approx 1.65\pm 0.19 \ \text{M}_{\odot} \ \text{yr}^{-1}$ in the Milky Way, \ \citealt{licquia2015}) for more than a few Gyr. Therefore, inflows of gas are necessary to prevent the quenching of star formation \citep[e.g.][]{fraternali2014}. The necessity of gas accretion is also required by galactic chemical evolution models \citep[e.g.][]{spitoni2021}: an example is the so-called G-dwarf problem \citep{van1962}, that is a deficit of metal-poor stars in the solar neighbourhood. This suggests the presence of an evolution for star-forming galaxies very different from total isolation. The preferred solution to this problem is to require a continuous accretion of low metallicity gas ($\sim 0.1 \ Z_{\odot}$, \citealt{larson1972, tosi1988, matteucci1989, colavitti2008}) that fuels star formation \citep{chiappini2001}, this can self-regulate, producing a gas close to solar metallicity and therefore what seems to be a deficit of metal-poor G-dwarf stars would be in reality an excess of solar metallicity G-dwarf stars. 

Gas accretion plays a primary role in the evolution of galaxies, but it remains a phenomenon that is very complex to model. Gas can be accreted from the intergalactic and circumgalactic media and this can occur generally with two possible scenarios: (i) a hot accretion mode, in which the gas falling into the potential well of the dark matter halo is shock heated to its virial temperature when it reaches the centre \citep{rees1977cooling,silk1977fragmentation,white1978core}, and (ii) a cold accretion mode, in which the gas can reach the centre of the halo still in a cold form \citep{birnboim2003virial,kerevs2005galaxies} because the post-shock cooling time becomes shorter than the dynamical time and the gas is not heated efficiently. A Milky Way-like galaxy ($M_{\text{h}} \sim 10^{12}$ M$_{\odot}$) typically reaches the critical mass necessary for the hot accretion mode at $z \sim 1 - 2$ \citep[e.g.][]{kerevs2009galaxies, huscher2021}, therefore it will form a gas reservoir (the so-called galactic corona) at the virial temperature around that time (although recent observations indicate the possible presence of a super-virial gaseous component in the Milky Way corona, \citealt{gupta2021}).

Galactic coronae are thought to be an essential component for the evolution of Milky Way-like galaxies and their presence is corroborated by different pieces of evidence. These gaseous haloes have a temperature higher than 0.1 keV (i.e. around the virial temperature of the halo), thus, the hot gas is expected to thermally emit X-ray photons through bremsstrahlung emission. In recent times many observations have been attempted to find and study the properties of these hot coronae around spiral galaxies, but, due to their temperature range and the low-densities, a large fraction of the thermal emission falls under the Galactic absorption threshold (0.2-0.3 keV) making this component very faint and highly contaminated by the X-ray background. For these reasons these observations failed to find extended X-ray emissions for years \citep[e.g.][]{li2008chandra,rasmussen2009hot,li2011x,bogdan2015hot}. Observations taken with \textit{Chandra}, \textit{XMM-Newton} and \textit{ROSAT} around massive ($\sim 10^{13}$ M$_{\odot}$) spiral galaxies, finally succeded in finding a visible X-ray emission of the corona in NGC 1961 \citep{anderson2011}, UGC 12591 \citep{dai2012}, NGC 266 \citep{bogdan2013}, NGC 6753 \citep{bogdan2017}, 2MASX J23453268-0449256 \citep{walker2015,mirakhor2021}. These observations suggest that the galactic coronae of these massive spiral galaxies have similar physical properties, such as masses, temperature profiles and metallicities. Instead, observations of galactic coronae around Milky Way-like galaxies are still extremely difficult, although it has recently been observed for the first time an extended X-ray emission in the L$^{\star}$ galaxy NGC 3221 \citep{dai2019,dai2020} and around the Large Magellanic Cloud \citep{krishnarao2022}. 

Despite the physical properties that make its direct detection very difficult, many attempts to constrain the density profile of the Milky Way corona have been carried out. A possible approach to estimate the density of the corona is to investigate the process of ram-pressure stripping of dwarf spheroidal galaxies inside the Milky Way halo, combining simulations of satellites orbiting around the Milky Way with observations \citep{grcevich2009,gatto2013,salem2015,troitsky2017, putman2021, martynenko2022}. Also, observations of the dispersion measure of pulsars can be used \citep{anderson2010}. Furthermore, another observational approach to study the coronal gas lies in observations of absorption lines at low impact parameters at redshift zero in the spectra of quasars \citep{yao2012,gupta2012,miller2013,troitsky2017,kaaret2020}. This has been done for L$^{\star}$ galaxies in the COS halo survey \citep{tumlinson2011}, which found out and confirmed the multiphase nature of the circumgalactic medium studying different physical properties of this gas, such as density, temperature, and metallicity. In this way it is possible to analyze different phases of the diffuse gas that surrounds L$^{\star}$ galaxies, spanning from neutral to highly ionized gas, mainly through O{\sc vii} and O{\sc viii} lines \citep[e.g.][]{werk2014, borthakur2016, werk2016}. A complementary approach is used by the survey AMIGA \citep{lehner2020}, that studied more in detail the circumgalactic medium of M31 using quasar sightlines at different impact parameters.

In the future, forthcoming missions (like \textit{XRISM} or \textit{Athena}) will be essential to study and to obtain detailed features for this component in spiral galaxies. That will hopefully enable us to simulate these galaxies in a more realistic way and to better understand how they evolve.

In the last years much progress has been made in our comprehension of how galaxies form and evolve and the use of hydrodynamical simulations \citep[see][for a review]{vogelsberger2020Review} has been an integral part of this success. Reproducing the observational properties of spiral galaxies has been particularly challenging. However, thanks to the implementation of effective stellar feedback processes and to their more accurate modelling \citep[e.g.][]{hopkins2011,tasker2011,wise2012,agertz2013,kannan2014phot,roskar2014}, simulations were able to generate realistic late-type galaxies \citep[e.g.][]{brook2012,aumer2013,marinacci2014,wang2015,colin2016,grand2017auriga, hopkins2018}.

In these simulations, stellar feedback acts as a source of negative feedback, it heats the gas eventually ejecting it from the disc, thus removing material to form new stars. However, the same process can act also as positive feedback. A possible mechanism that could be triggered is represented by the interaction between the cold, high-metallicity gas ejected from the disc and the hot, metal-poor corona. Star-forming galaxies can eject gas in the form of galactic fountains \citep{shapiro1976, fraternali2006} or galactic winds \citep{oppenheimer2010}; if this gas efficiently mixes with the corona it decreases its cooling time, forming an intermediate temperature and metallicity gas phase. Within this gas phase, some of the coronal gas can condensate because of its reduced cooling time, increasing the mass of the cold gas that then rains down on the disc, supplying it with fresh fuel for star formation. This phenomenon has been studied with parsec-scale simulations of the interface region between the corona and the cold gas clouds \citep{marinacci2010,marinacci2011,armillotta2016} and also in more recent works. For instance, \citet{hobbs2020} studied the effect of this positive feedback on the star formation of the galaxy with 3D simulations of a portion of a disc, \citet{dutta2022} developed analytic solutions for the radiative cooling flow around small cold gas clumps that are present in the CGM and \citet{gronke2022} studied the survival of a gas cloud in a turbulent medium. However, in these studies the nature of the interaction is captured but not inserted in a realistic galaxy model. Furthermore, the balance between the negative and positive nature of stellar feedback is actually not well understood. In particular, the interaction between the cold disc of the galaxy and the hot galactic corona has not been analysed in detail in galactic-scale simulations.

This paper is focused on the study of the gas circulation between the disc and the corona of star-forming galaxies like the Milky Way. To analyse this circulation and its implications on the properties of the corona and of the gas that is subsequently accreted onto the disc, we make use of high-resolution ($\sim 10^4 \ \text{M}_{\odot}$) hydrodynamical $N$-body simulations of an isolated Milky Way-like galaxy, with the inclusion of an observationally-motivated hot corona around the galaxy. After sampling the galaxy initial conditions, the simulations are evolved with the \smuggle\ model \citep{marinacci2019simulating}, an explicit ISM and stellar feedback model that is part of the moving-mesh code \arepo\ \citep{springel2010pur, weinberger2020arepo}. So far, the \smuggle\ model has been tested on isolated galaxies in the absence of a hot corona \citep[e.g.][]{kannan2020, hui2020, kannan2021, jahn2021, burger2022a, li2022, angus2022, smith2021, sivasankaran2022}, which is usually not included in this type of hydrodynamical $N$-body simulations. The calculations that we carry out in this paper are the first that include such an important component for galactic evolution in the \smuggle\ framework.

The paper is structured as follows: in Section \ref{num_met} we describe the numerical methods used in this work, in particular the \smuggle\ model. Section \ref{ics} details how we have created the initial conditions for our simulations. In Section \ref{res} we describe the structure of the ISM and the galactic corona that is generated in the simulations. In Section \ref{starform_circ} we present the main results of this work. In Section \ref{caveats} we compare our results to previous works and we summarize them in Section \ref{conclusions}.

\section{Numerical Methods}\label{num_met}
\subsection{The moving-mesh code \arepo}\label{simcode}
The simulations in this work are performed with the hydrodynamical $N$-body code \arepo\ \citep{springel2010pur,weinberger2020arepo}. \arepo\ is a moving-mesh code that uses a Tree-PM method coupled with a leapfrog scheme to solve for gravitational dynamics and a finite volume solver on an unstructured Voronoi mesh for hydrodynamics. The fluxes at the interface of each cell are computed with an exact Riemann solver and the conservative update of the variables at each time-step is done with a second order Runge-Kutta scheme \citep{pakmor2016}. The presence of a mesh that is free to move with the fluid mitigates the problems that affect Eulerian and Lagrangian codes, combining the high spatial adaptivity of Lagrangian codes and the ability of accurately resolve discontinuities of Eulerian codes. One of the main features of this code is its explicit Galilean-invariance. Without it, the accuracy of the hydrodynamical schemes would be reduced in the presence of large bulk velocities. This is particularly important given the fact that supersonic flows are common in the majority of astrophysical phenomena. Thanks to its flexibility, in the last years \arepo\ has been successfully used in computational astrophysics in a vast number of topics, for instance in state-of-the-art cosmological simulations of galaxy formation and evolution, such as Illustris \citep{genel2014introducing,vogelsberger2014introducing,2014vogelsbergerNature}, IllustrisTNG \citep{Naiman_2018,Nelson_2017,Pillepich_2017,Marinacci_2018,Springel_2017} and Auriga \citep{marinacci2017properties,grand2017auriga}, but also in smaller scale simulations of different astrophysical objects, like isolated galaxies \citep[e.g.][]{jacob2018dependence,pascale2021off}, active galactic nuclei \citep[e.g.][]{bourne2021,ehlert2022}, molecular filaments and clouds in galaxies \citep[e.g.][]{smith2014co,galeano2022}, type Ia supernovae \citep[e.g.][]{pakmor2013helium,pakmor2022} and protoplanetary discs \citep[e.g.][]{munoz2014planet}.

\subsection{The \smuggle\ model}\label{smuggle}

We coupled \arepo\ with the \textbf{S}tars and \textbf{MU}ltiphase \textbf{G}as in \textbf{G}a\textbf{L}axi\textbf{E}s model (\textbf{\smuggle}; \citealt{marinacci2019simulating}), an explicit stellar feedback and ISM medium model. \smuggle\ is a sub-resolution model that accounts for the main phenomena that drive the evolution of galaxies: it describes the complex multi-phase structure of the ISM and self-consistently allows the galaxy to generate gaseous outflows. This model includes: cooling and heating of the gas, a stellar evolution model and stellar feedback. The cooling incorporates radiative emission from a primordial mixture of H and He and from metal lines \citep[implemented as in][]{vogelsberger2013}. In addition, the \smuggle\ model accounts for low-temperature metal lines, fine structure and molecular cooling processes \citep{hopkins2018}. These cooling channels are particularly important, as they allow the gas to reach very low temperatures ($T \approx 10 $ K) and, consequently, to form stars. 
Also, \smuggle\ considers photoelectric \citep{wolfire2003} and cosmic ray \citep{guo2008} gas heating. However, the model does not include magnetic fields and an accurate cosmic rays dynamics. We briefly describe the potential impact of these phenomena in Section \ref{conclusions}.
The \smuggle\ model contains also a star formation module, for which a gas cell turns into a star particle if (i) the density is higher than a density threshold $\rho_{\text{th}} = 100$ cm$^{-3}$ and if (ii) a condition that ensures whether the region is gravitationally bound is satisfied (see \citealt{marinacci2019simulating} for details). The star particles are complemented with the stellar evolution model from \citet{vogelsberger2013}; each star particle evolves in time and loses mass and metals depending on its evolutionary phase. 

The newly formed stars exert feedback on the ISM, thus injecting energy and momentum in the surrounding medium. This can happen through three main feedback channels: 

(i) \textit{supernova (SN) feedback}, including type II and type Ia SNe: the number of type II SNe at each time-step is obtained integrating the initial mass function, that is chosen as a \citet{chabrier2003galactic}, assuming that only stars between $M_{\text{SNII,min}}=8 \ M_{\odot}$ and $M_{\text{SNII,max}}=100 \ M_{\odot}$ can explode as type II SNe. The rate of type Ia SNe is parametrized with a delay time distribution \citep[e.g.][]{mannucci2006}. From these rates the consequent injection of energy, momentum, mass and metals into the ISM is computed. Due to the limited resolution we cannot fully resolve the cooling radius of the SN remnants and the consequent generation of momentum during the Sedov-Taylor phase of the SN. This momentum generation is not negligible and is accounted for in the model \citep[see][]{marinacci2019simulating};

(ii) \textit{radiative feedback}, that includes photoionization from young massive stars and radiation pressure: the photoionization is treated with a probabilistic approach in which the probability of a gas cell of being photoionized is defined by the ionizing photon rate. When a cell of gas is considered photoionized its temperature is set to $T_{\text{phot}}=1.7 \times 10^4$ K and its cooling is disabled for a time equal to the star particle time-step. The radiation emitted from stars generates pressure and is therefore a source of momentum. This momentum is injected in the gas cells surrounding the star particle;

(iii) \textit{stellar wind feedback}, the \smuggle\ model accounts for stellar winds generated by two classes of stars: young and massive type OB stars and asymptotic giant branch (AGB) stars. These provide two channels of feedback that act at different times \citep[e.g.][]{matzner2002, krumholz2009}. At first, the model computes the mass loss from the two types of stars. From that, it injects the associated energy and momentum into the medium that surrounds the star particle. 

Owing to the aforementioned physical processes, \smuggle\ is able to: (i) generate an ISM with different cohexisting gas phases: cold, warm and hot gas in pressure equilibrium that roughly matches the observed densities and fractions observed in the Milky Way \citep{ferriere2001}; (ii) self-consistently launch gaseous outflows outside the disc of the galaxy; (iii) reproduce the observed Kennicut-Schmidt relation. We refer the reader to \citet{marinacci2019simulating} for a more detailed description of the model and the implementation of the feedback channels.

\section{Initial conditions}\label{ics}
\label{sec:maths} 

In this Section we describe the initial conditions (ICs) of the Milky Way-like multi-component galaxy model that we study in this work by means of hydrodynamical $N$-body simulations with the moving-mesh code \arepo. The ICs of the galaxy have been generated according to the approach adopted in \citet[][see also \citealt{hernquist1993n} and \citealt{springel2000modelling}]{springel2005modelling}. This method enables construction of 
a multi-component galaxy with all the components in approximate equilibrium. These components include: a static dark matter halo and a live stellar bulge both of which follow a Hernquist profile (Section \ref{dark_bulge}), an exponential thick stellar disc, an exponential thick gaseous disc (Section \ref{stell_disc}) and a galactic hot corona (Section \ref{gal_corona}).

\subsection{Dark matter halo and bulge}\label{dark_bulge}
The dark matter halo and the bulge are modeled using a spherical \citet{hernquist1990analytical} profile

\begin{equation}\label{profil_hern}
	\rho_{\text{comp}}(r)=\frac{M_{\text{comp}}}{2\pi} \frac{a_{\text{comp}}}{r(r+a_{\text{comp}})^3},
\end{equation}
where $\text{comp}=\text{dm}$ for the dark matter halo and $\text{comp}=\text{b}$ for the bulge. $a_{\text{comp}}$ and $M_{\text{comp}}$ indicate the scale radius and the total mass of each component, respectively, and $r=\sqrt{x^2+y^2+z^2}$ is the spherical radius. The dark matter halo is modeled as a static gravitational field for efficiency reasons.

Regarding the velocity distribution, since the total potential is axisymmetric when we include the discs contribution (see Section \ref{stell_disc}), we assume that the distribution function of bulge depends only on the energy and on the $z$-component of the angular momentum $L_z$. In this configuration the mixed second order velocity momenta and the first momenta in radial and vertical directions are null $\langle v_R v_{\phi}  \rangle = \langle v_R v_z \rangle =  \langle v_{\phi} v_z \rangle=0$, $\langle v_R   \rangle = \langle v_z \rangle=0$ \citep{binney2008}.
The particle velocities for the bulge are derived from the Jeans equations using the so-called \textit{Gaussian approximation}: we assume that, at each point $r$, the velocities ($v_R$, $v_z$, $v_{\phi}$) in cylindrical coordinates follow a triaxial Gaussian distribution, from which they are randomly sampled. With this method we are not computing the exact equilibrium distribution, but this approximation is good enough for the purpose of this work.
 The non-vanishing second-order momenta can be obtained from the Jeans equations as:
\begin{equation}\label{vel1}
	\langle v_z^2 \rangle=\langle v_R^2 \rangle=\frac{1}{\rho_{\text{b}}} \int_{z}^{\infty} \rho_{\text{b}}(z,R)\frac{\partial \Phi}{\partial z} \, \text{d}z \ ,
\end{equation}

\begin{equation}\label{vel2}
	\langle v_{\phi}^2 \rangle=\langle v_R^2 \rangle + \frac{R}{\rho_{\text{b}}} \frac{\partial (\rho_{\text{b}} \langle v_R^2 \rangle )}{\partial R} + v_c^2,
\end{equation}
where $v_c= R$ $\partial \phi / \partial R$ is the circular velocity, $\rho_{\text{b}}$ is the density of the bulge, $\langle v_{\phi}^2 \rangle$ is the azimuthal second order velocity moment and $\langle v_R^2 \rangle$ is the radial second order velocity moment, for which we have omitted the subscript "$\text{b}$" for simplicity. $\Phi$ is the total gravitational potential, which includes the contribution of all galaxy components. 
We assume that the bulge has no net rotation.

\subsection{Stellar and gaseous discs}\label{stell_disc}
Observationally, it has been shown that disc galaxies have radial exponential profiles \citep{freeman1970,portinari2010}; for this reason the stellar and the gaseous discs have been modeled with exponential surface density profiles

\begin{equation}\label{star_surf}
 	\Sigma_{\text{comp}}(R)=\frac{M_{\text{comp}}}{2\pi h_{\text{comp}}^2}\text{exp}(-R/h_{\text{comp}}) \ ,
 	\end{equation}
 where $\text{comp}=\star$ for stars and $\text{comp}=\text{g}$ for gas. $h_{\text{comp}}$ is the scale length and $M_{\text{comp}}$ is the mass of the stellar/gaseous disc. 	
\begin{table*}
	\centering
	\caption{\textit{Structural parameters of the Milky Way-like galaxy simulated in this work}. From left to right: circular velocity of the halo at $r_{200}$ ($v_{200}$); dark matter halo mass ($M_{\text{dm}}$); dark matter halo scale length ($r_{\text{s}}$); bulge mass ($M_{\text{b}}$); bulge scale length ($a$); stellar disc mass ($M_{\star}$); stellar disc scale length ($h_{\star}$); stellar disc scale height ($z_0$); gaseous disc mass ($M_{\text{g}}$); gaseous disc scale length ($h_{\text{g}}$); mass of the corona ($M_{\text{cor}}$) computed within the virial radius $r_{\text{vir}}$; corona core radius ($r_{\text{c}}$).}\label{tabella_param}
	\begin{tabular}[t]{cccccccccccc}
	
		\toprule
		$v_{200}$&$M_{\text{dm}}$&$r_{\text{s}}$&$M_{\text{b}}$ & a & $M_{\star}$&$h_{\star}$& $z_0$ &$M_{\text{g}}$ & $h_{\text{g}}$ & $M_{\text{cor}}$&$r_{\text{c}}$\\

		[km s$^{-1}$]&$[M_{\odot}]$&[kpc]&$[M_{\odot}]$&[kpc]&$[M_{\odot}]$&[kpc] & [pc] & [$M_{\odot}$] & [kpc] & [$M_{\odot}$]&[kpc]\\
		\bottomrule
	    \rule{0pt}{3ex}  
		$169$ & $1.53\times 10^{12}$&36.46&$1.5\times 10^{10}$&1.3&$4.73\times 10^{10}$&3.8 & 380 & $9\times 10^{9}$ & 7.6 & $3\times 10^{10}$ &8\\
		
		\bottomrule
	\end{tabular}
\end{table*}%
Vertically the stellar disc follows a sech$^2$ profile, that corresponds to isothermal sheets perpendicular to the disc plane \citep{binney2008}
\begin{equation}\label{sec}
	\rho_{\star}(R,z)= \frac{\Sigma_{\star}(R)}{2 z_0 } \text{sech}^2\biggl(\frac{z}{z_0} \biggr),
	\end{equation}
where $z_0$ is a free parameter that determines the scale height of the stellar disc.

\begin{figure}
	\includegraphics[width=\columnwidth]{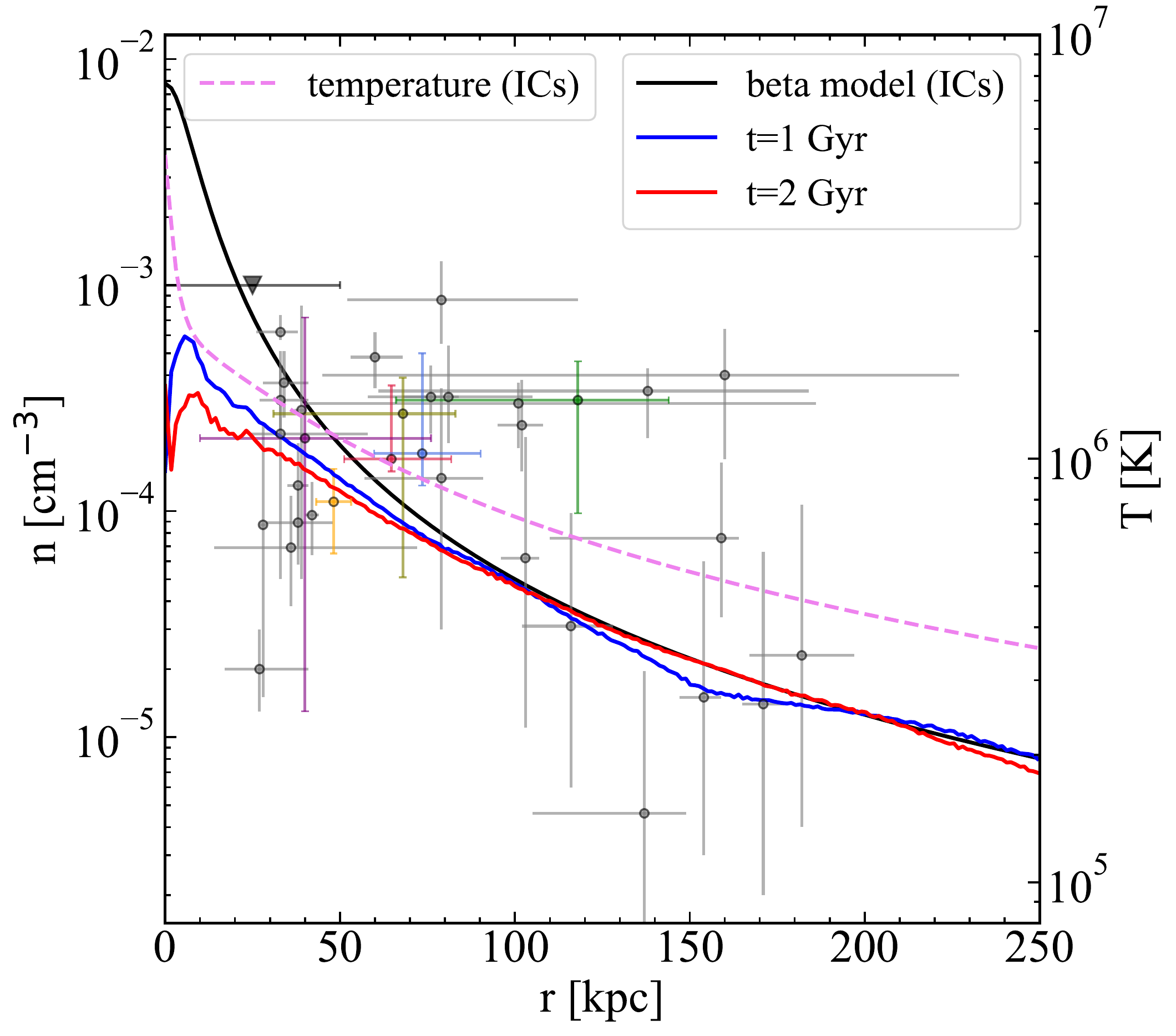}
    \vspace*{-5mm}\caption{Galactic corona number density profile (black solid line) of the fiducial simulation {\sc BM\_r} (see Table \ref{model_var}) compared with several observational constraints of the Milky Way (points with errorbars): ram-pressure stripping from the dwarf spheroidal galaxies Fornax, Ursa Minor and Sculptor (green, purple and olive points, \citealt{grcevich2009}), Carina and Sextans (red and blue points, \citealt{gatto2013}), the Large Magellanic Cloud (orange point, \citealt{salem2015}) and also more recent constraints obtained from several dwarf galaxies (grey points, \citealt{putman2021}); and Pulsar dispersion measures (black upper limit, \citealt{anderson2010}). The black line represents the initial number density corona profile (i.e. a $\beta$-model), while the blue and the red solid lines represent the same profile at 1 Gyr and 2 Gyr after the start of the simulation respectively. The purple dashed line is the initial temperature of the coronal gas.}
    \label{hqbm}
\end{figure}

Instead, for the gaseous disc the vertical distribution is determined by the vertical hydrostatic equilibrium using an iterative method. For a given potential the hydrostatic equilibrium equation is solved obtaining $\rho_g$, imposing that
\begin{equation}\label{sigma_int}
	\Sigma_g(R)=\int_{-\infty}^{+\infty} \rho_g(R,z) \, \text{d}z,
	\end{equation}
where $\Sigma_g$ is the gas surface density in equation (\ref{star_surf}).
The scale length of the stellar disc $h_{\star}$ is found iteratively (see \citealt{springel2005modelling} for more details) in order to obtain the best approximation for the disc angular momentum.

For the stellar disc we adopt a Gaussian approximation for the velocity distribution as well.
Observations suggest that $\langle v_R^2 \rangle$ is proportional to $\langle v_z^2 \rangle$ \citep{hernquist1993n, garma2021}, in particular we assumed $\sigma^2_R=\langle v_R^2 \rangle= \langle v_z^2 \rangle$. 
To obtain the mean streaming velocity we use the epiciclic approximation, valid for quasi-circular orbits in an axisymmetric potential.

For the gaseous disc, we assume that the only non null component is the azimuthal one, that can be obtained from the stationary equation of momentum conservation in the radial direction, this yields
\begin{equation}\label{vel4}
v_{\phi,\text{gas}}^2=R \biggl( \frac{\partial \phi}{\partial R} + \frac{1}{\rho_g}\frac{\partial P}{\partial R} \biggr).
\end{equation}

The gaseous components in the ICs are also characterized by a temperature profile and a metallicity distribution. The initial temperature in the disc is assumed to be constant at the value of $10^4$ K. However, when the galaxy is evolved in time, we expect the formation of a multiphase gas, with both very low ($\sim 10-100$ K) and very high temperatures ($\sim 10^6$ K), due to cooling and heating mechanisms and to the stellar feedback. The overall metallicity of the gas in the disc is set to $Z=Z_{\odot}$ for simplicity.

\subsection{Galactic corona}\label{gal_corona}
The properties of the hot galactic corona have been set using some of the observational constraints available for this component in the Milky Way \citep{grcevich2009,anderson2010,gatto2013,salem2015, putman2021}. We modeled the corona with a $\beta$-model \citep{cavaliere1978}, that is often used to fit the density profiles of the hot gas in galaxy clusters and that has been shown to fit well the corona density profile of the L$^*$ galaxy NGC 3221 in recent observations \citep[e.g.][]{dai2020}
\begin{equation}
    \rho_{\text{cor}}(r) = \rho_0 \biggl[1 + \biggl(\frac{r}{r_c}\biggr)^2\biggr]^{-\frac{3}{2} \beta}.
\end{equation}
The parameters for the $\beta$-model are taken following \citet{moster2011}. $\beta$ is set to $2/3$ and the core radius is set to $r_{\text{c}}=0.22r_s=8$ kpc.

In Figure \ref{hqbm} we show a comparison between the number density of the coronal gas (black solid line) of the fiducial simulation (Table \ref{model_var}) with observational estimates of the Milky-Way corona (points with errorbars). The model has a mass of $3\times 10^{10}$ M$_{\odot}$ inside the virial radius $R_{200}\approx240$ kpc and is well constrained by the observational points. The $\beta$-profile has a region of constant density (i.e. a core) in the centre, that depends on the parameter $r_{\text{c}}$: small changes in the inner region density can have important consequences on the cooling of this component, as the cooling time is extremely dependent on the gas density and temperature. We have also shown the evolution of the number density over time, in particular at 1 Gyr (blue solid line) and 2 Gyr (red solid line) after the beginning of the simulation. The galactic corona slowly accretes into the disc of the galaxy and the density is decreasing particularly in the central region (where it changes from $8\times10^{-3}$ to $3\times10^{-4}$ cm$^{-3}$). At large radii ($r>50$ kpc) the density is similar to the initial one. At both times the density profile is in good agreement with the observational points.

\begin{figure}
\centering
	\includegraphics[width=0.8\columnwidth]{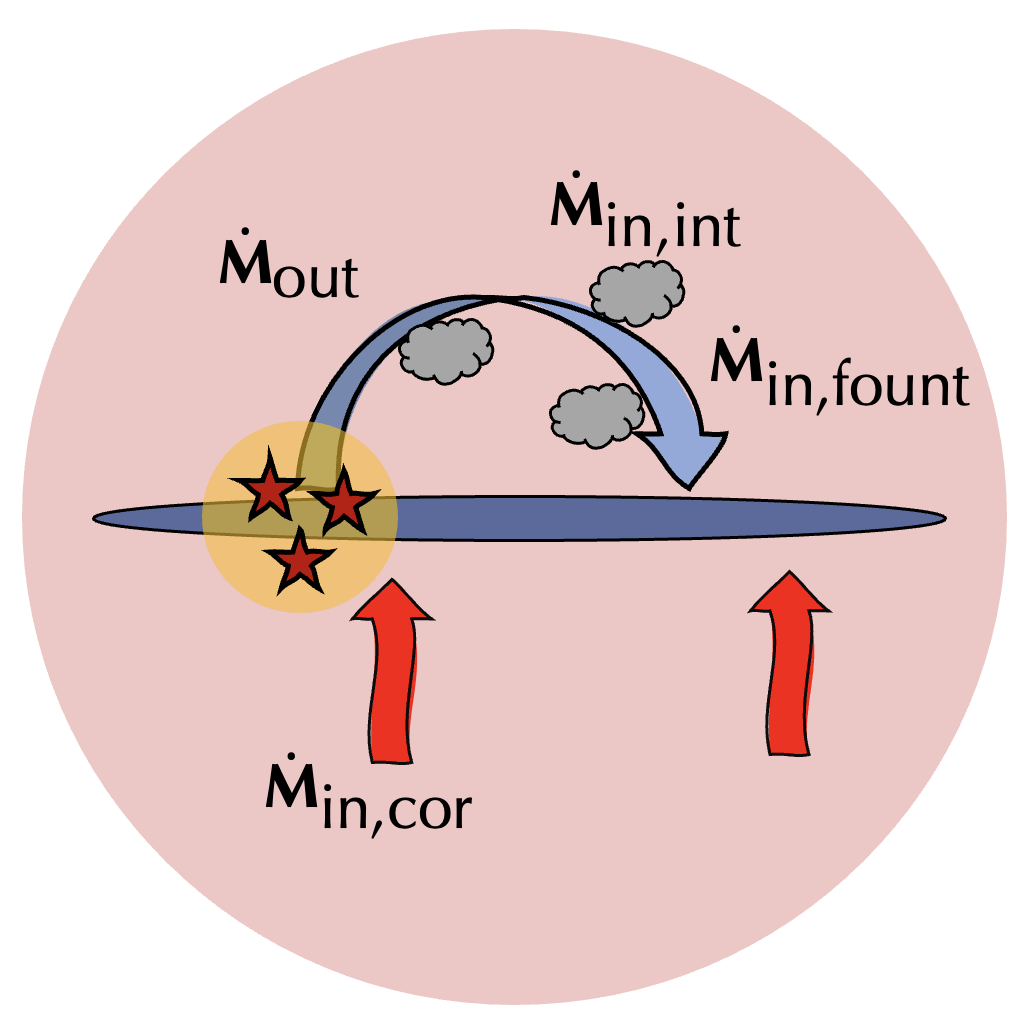}
	\caption{Cartoon scheme of the gas accretion/ejection processes in a Milky Way-like galaxy. The stellar feedback processes (i.e. SN, stellar winds and radiative feedback) can eject gas outside the galactic disc. The outflow rate ($\dot{M}_{\text{out}}$) is represented by the blue arrow, that schematize the trajectory of a single galactic fountain. This gas can eventually fall back onto the disc, giving a inflow rate coming from the galactic fountains ($\dot{M}_{\text{in, fount}}$). The interaction between the fountains and the corona (red halo) can create a mixture of gas, represented by the grey clouds, that can increase the gas accretion rate onto the disc ($\dot{M}_{\text{in, int}}$). The red arrows represent the inflow rate of gas accreted directly from the galactic hot corona ($\dot{M}_{\text{in, cor}}$). }
    \label{galaxy_scheme}
\end{figure}

\begin{figure*}
\centering
\subfloat{\includegraphics[width=0.28\textwidth]{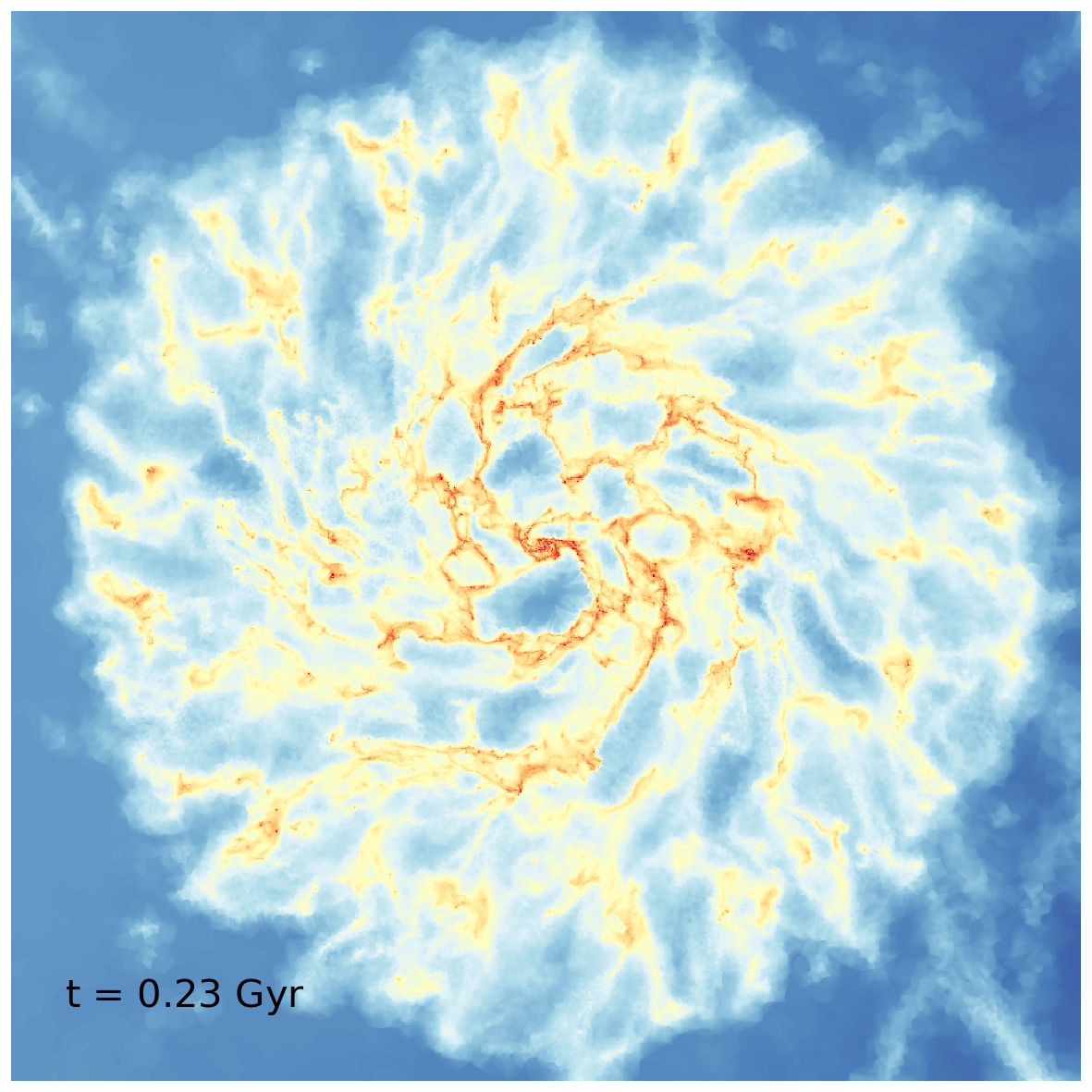}} 
\subfloat{\includegraphics[width=0.28\textwidth]{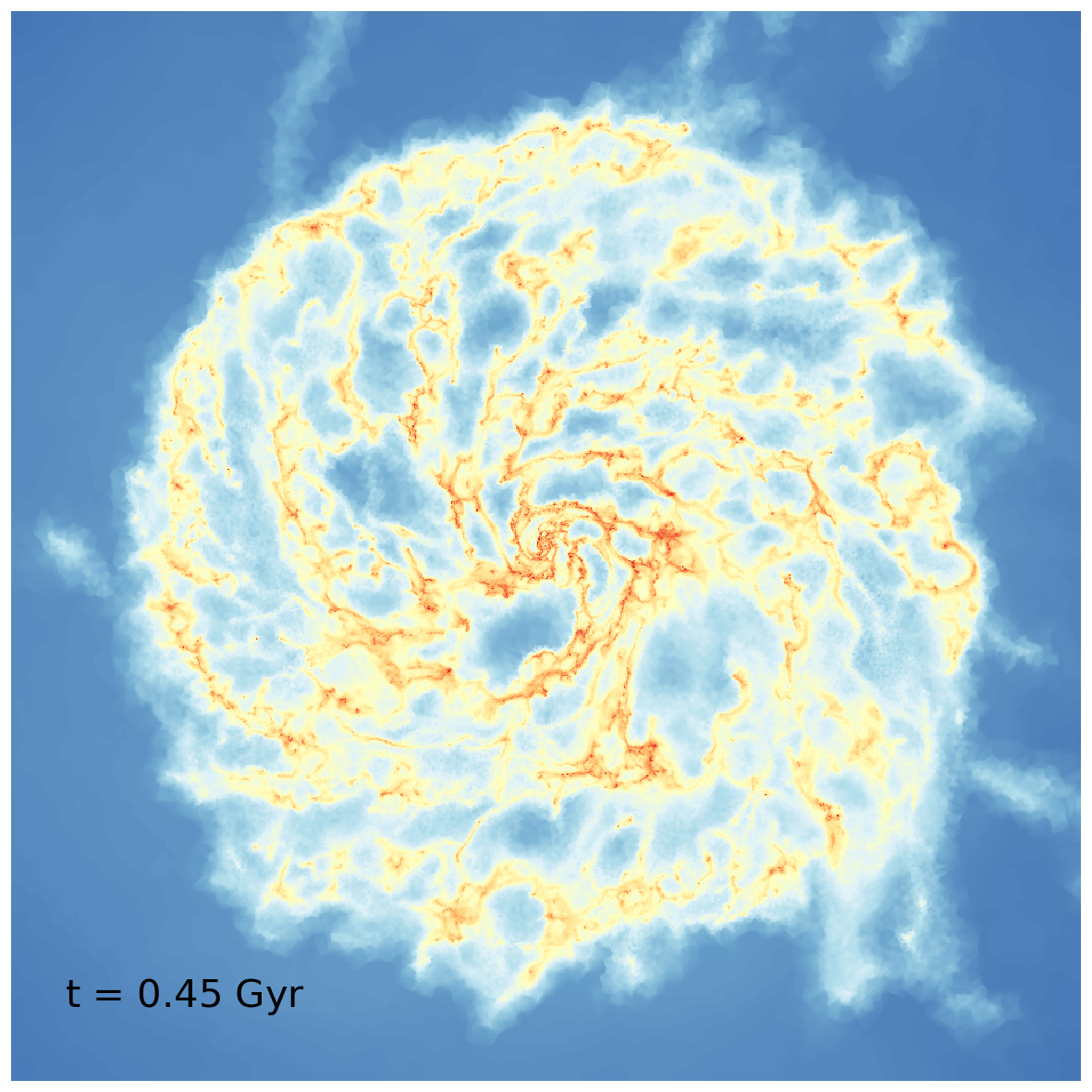}}
\subfloat{\includegraphics[width=0.28\textwidth]{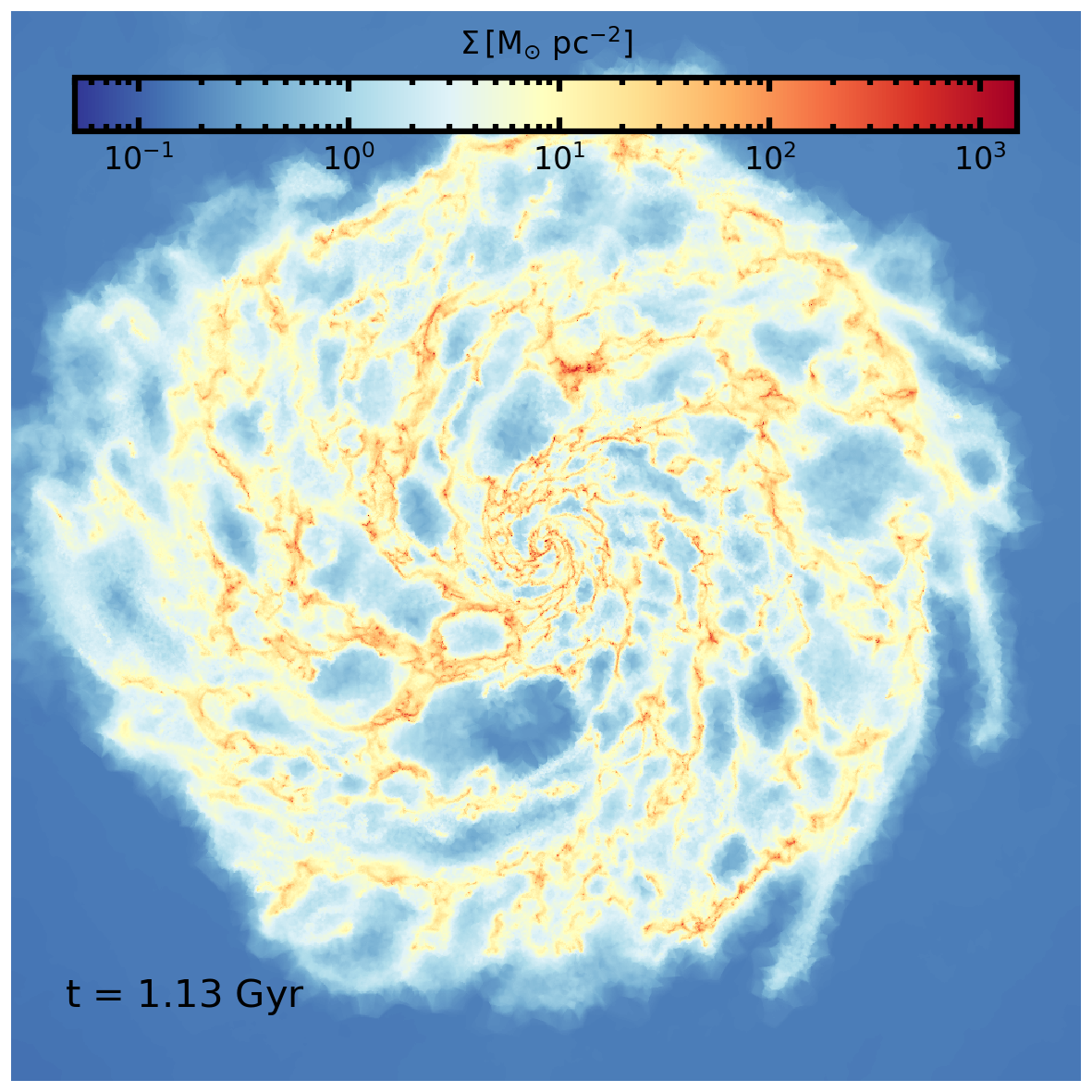}}\\[-3ex] 
\subfloat{\includegraphics[width=0.28\textwidth]{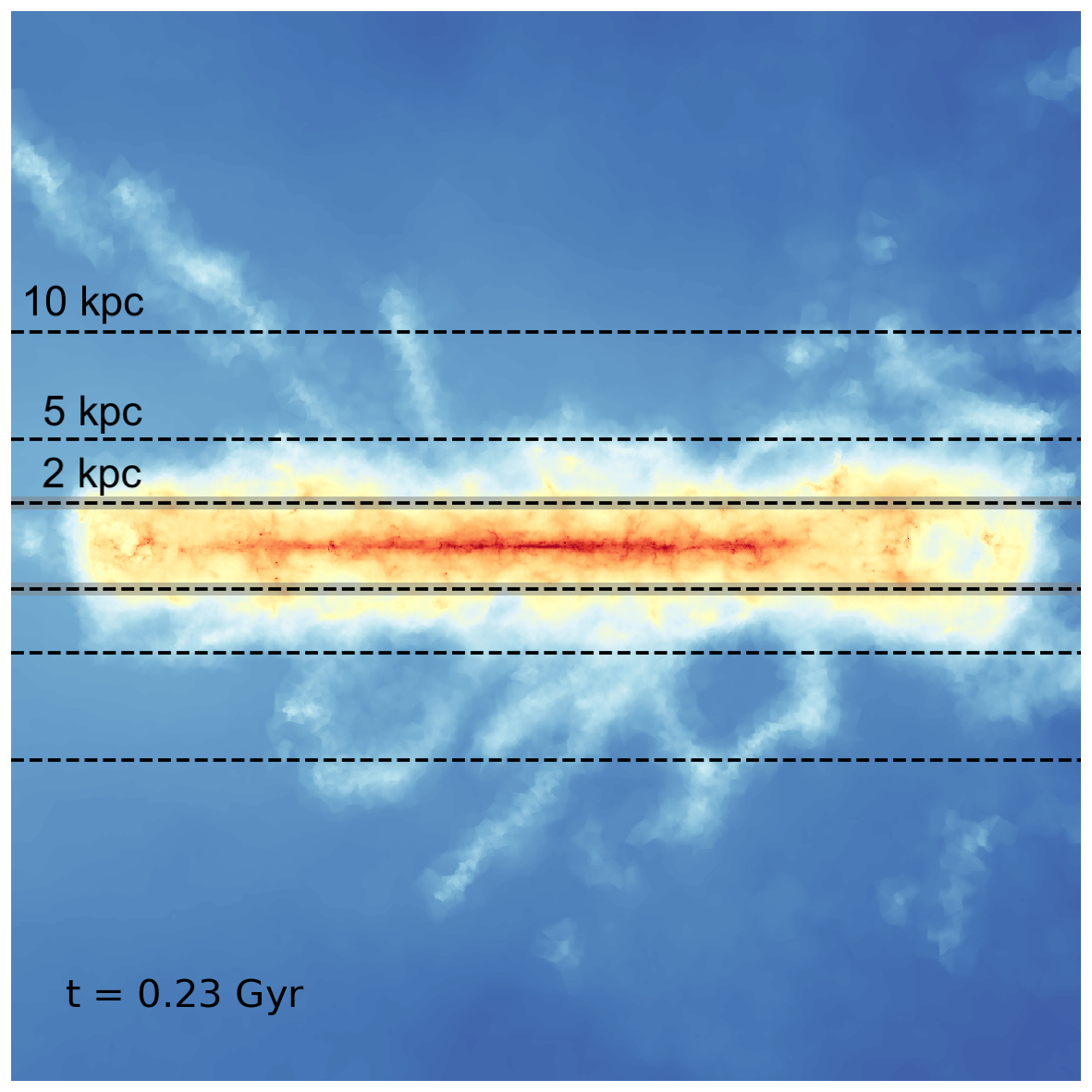}}  
\subfloat{\includegraphics[width=0.28\textwidth]{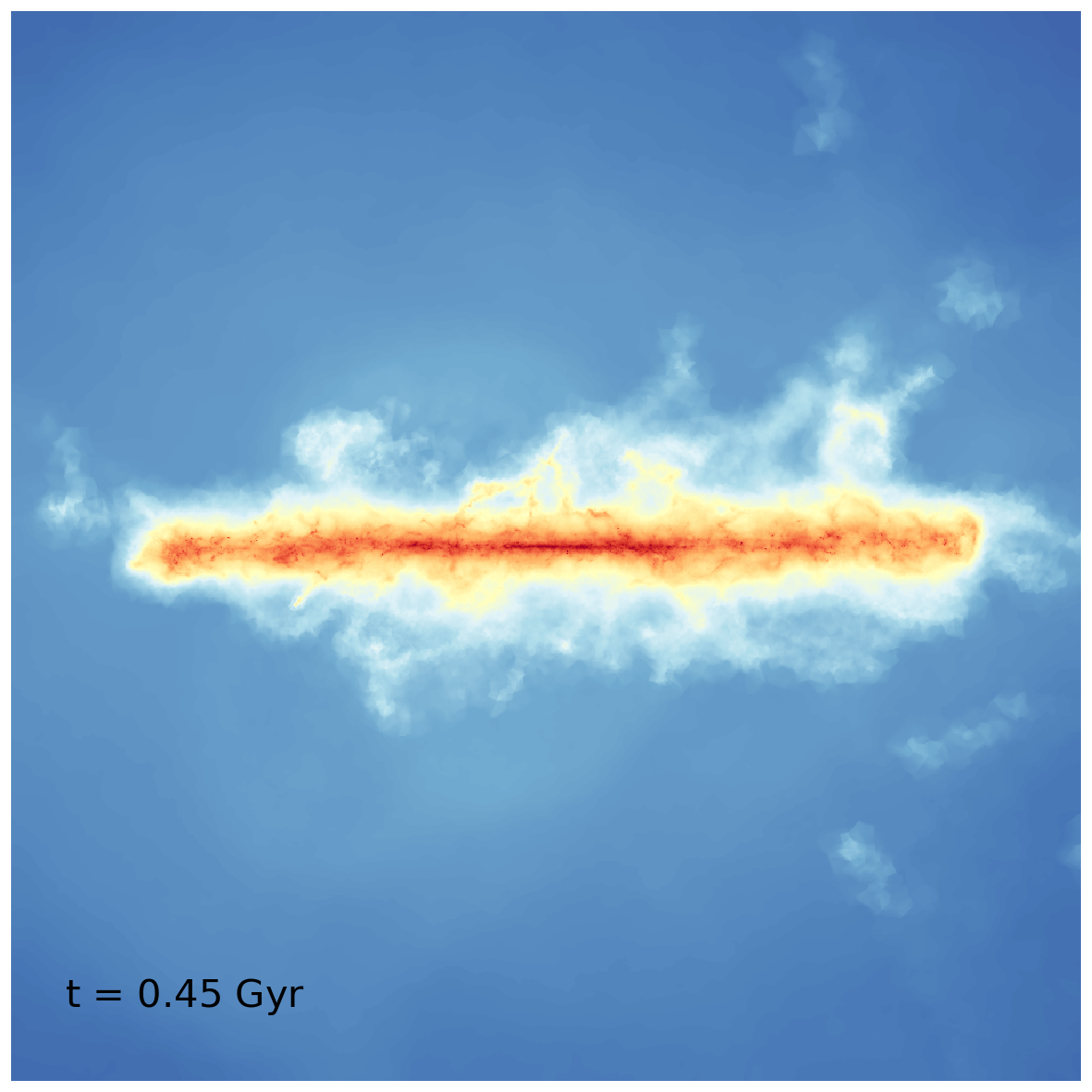}}
\subfloat{\includegraphics[width=0.28\textwidth]{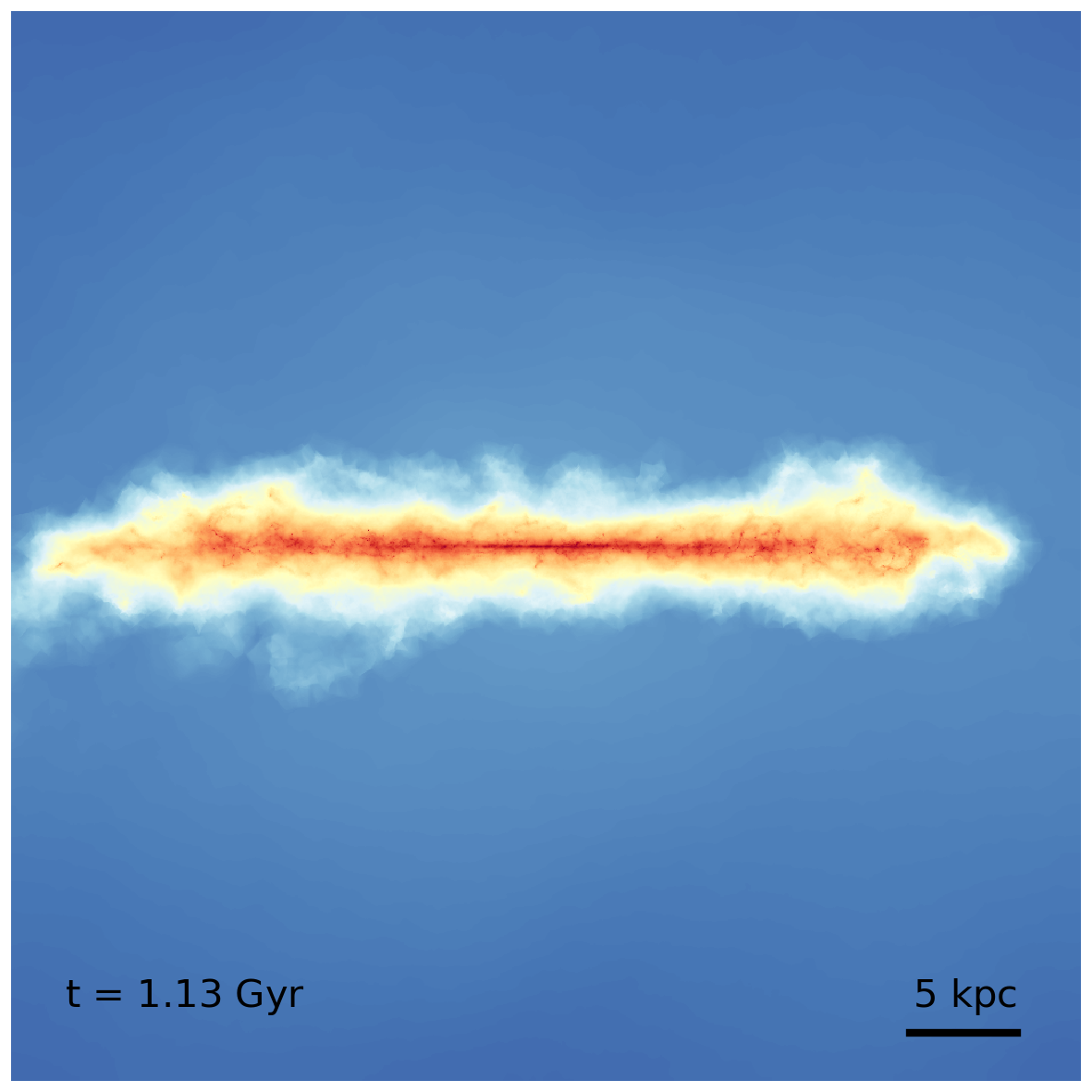}}\\[-3ex] 
\subfloat{\includegraphics[width=0.28\textwidth]{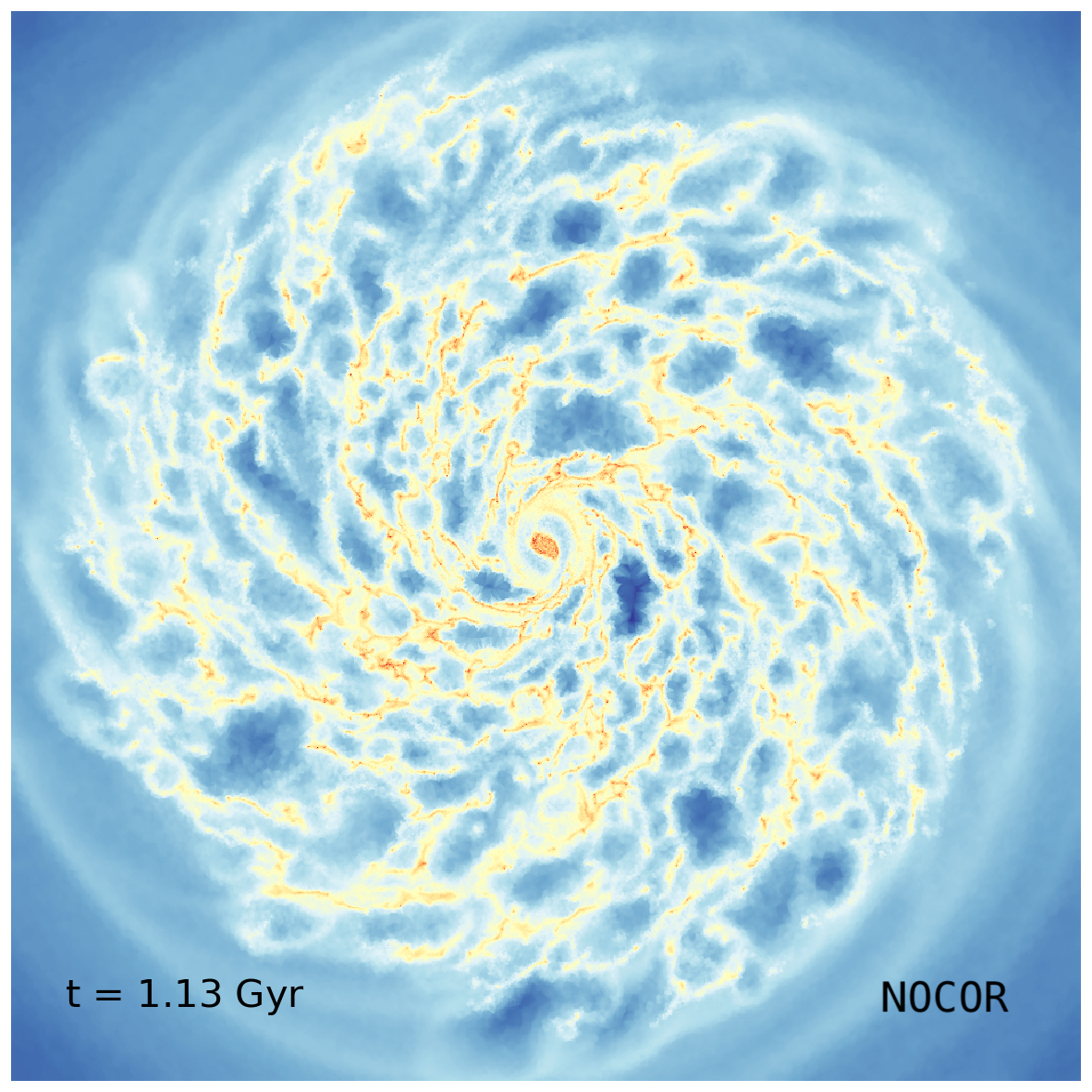}}
\subfloat{\includegraphics[width=0.28\textwidth]{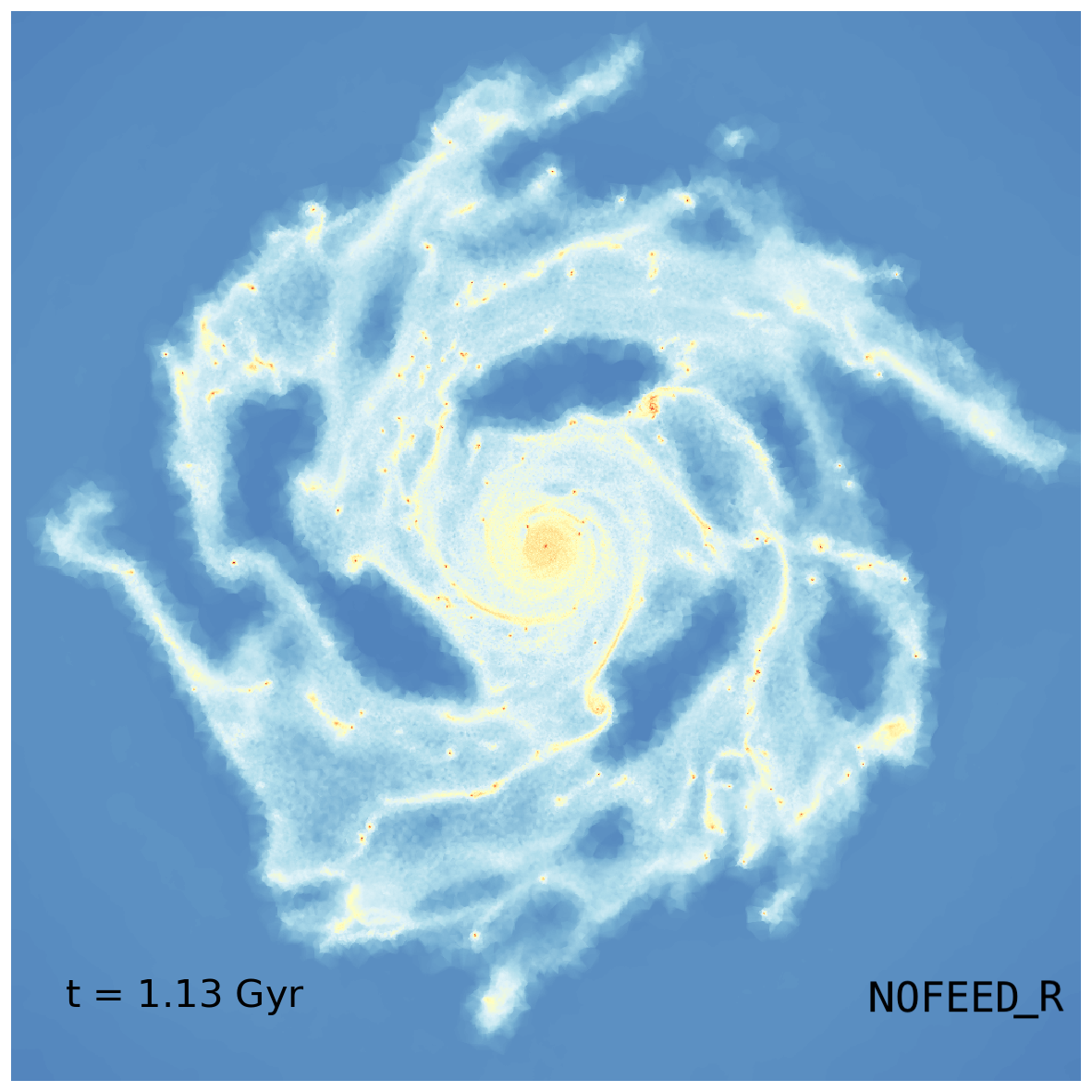}}
\subfloat{\includegraphics[width=0.28\textwidth]{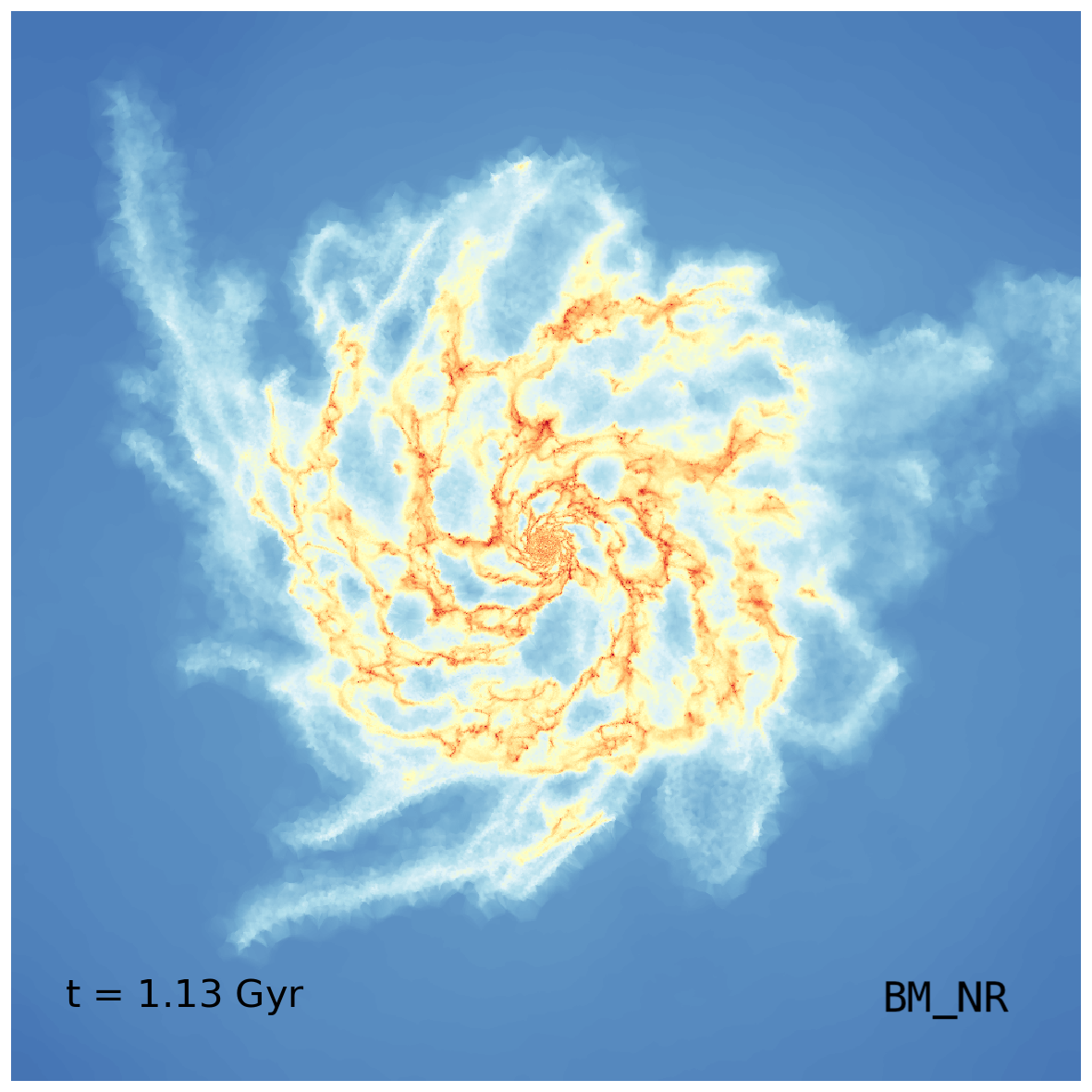}} \\[-3ex] 
\subfloat{\includegraphics[width=0.28\textwidth]{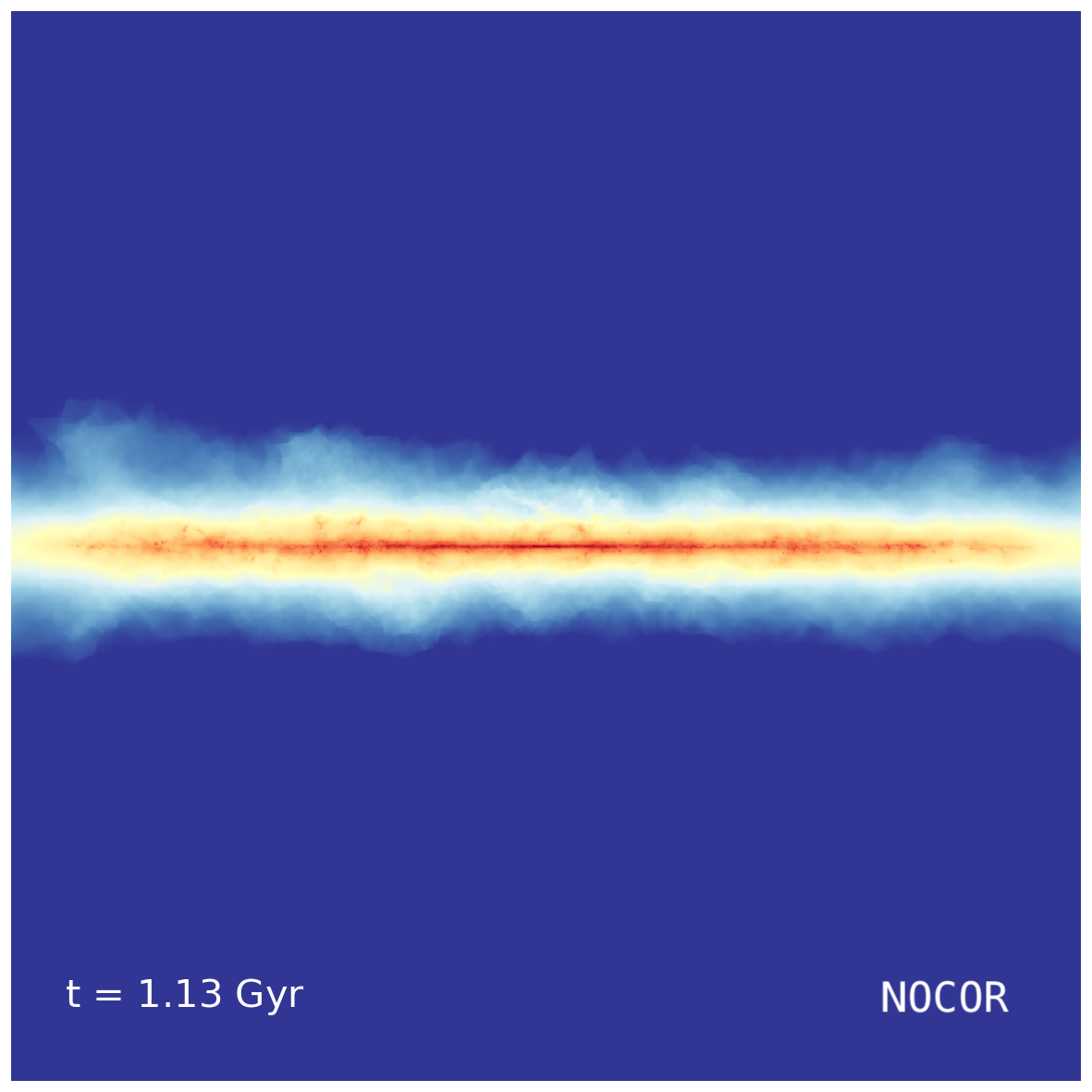}}  
\subfloat{\includegraphics[width=0.28\textwidth]{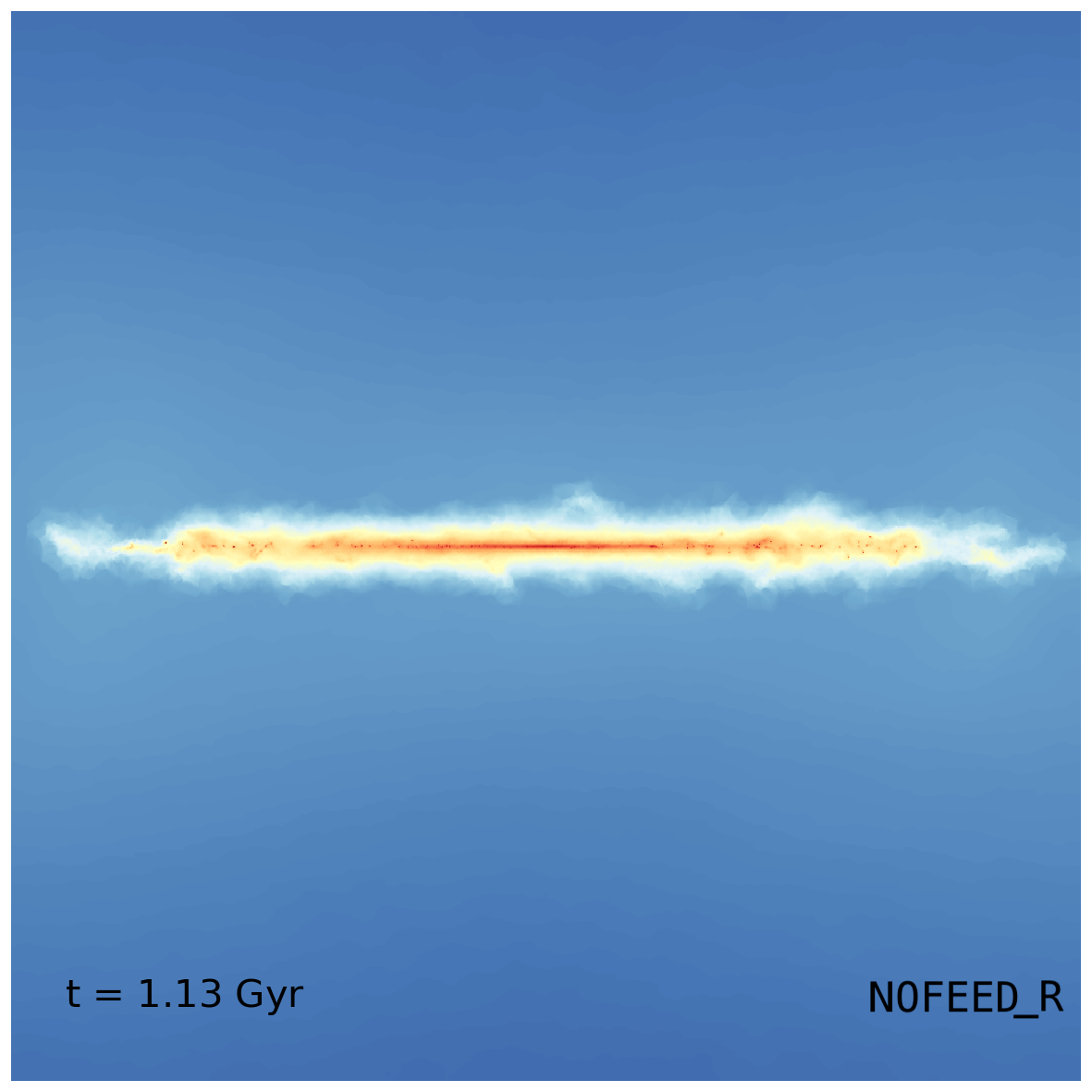}}
\subfloat{\includegraphics[width=0.28\textwidth]{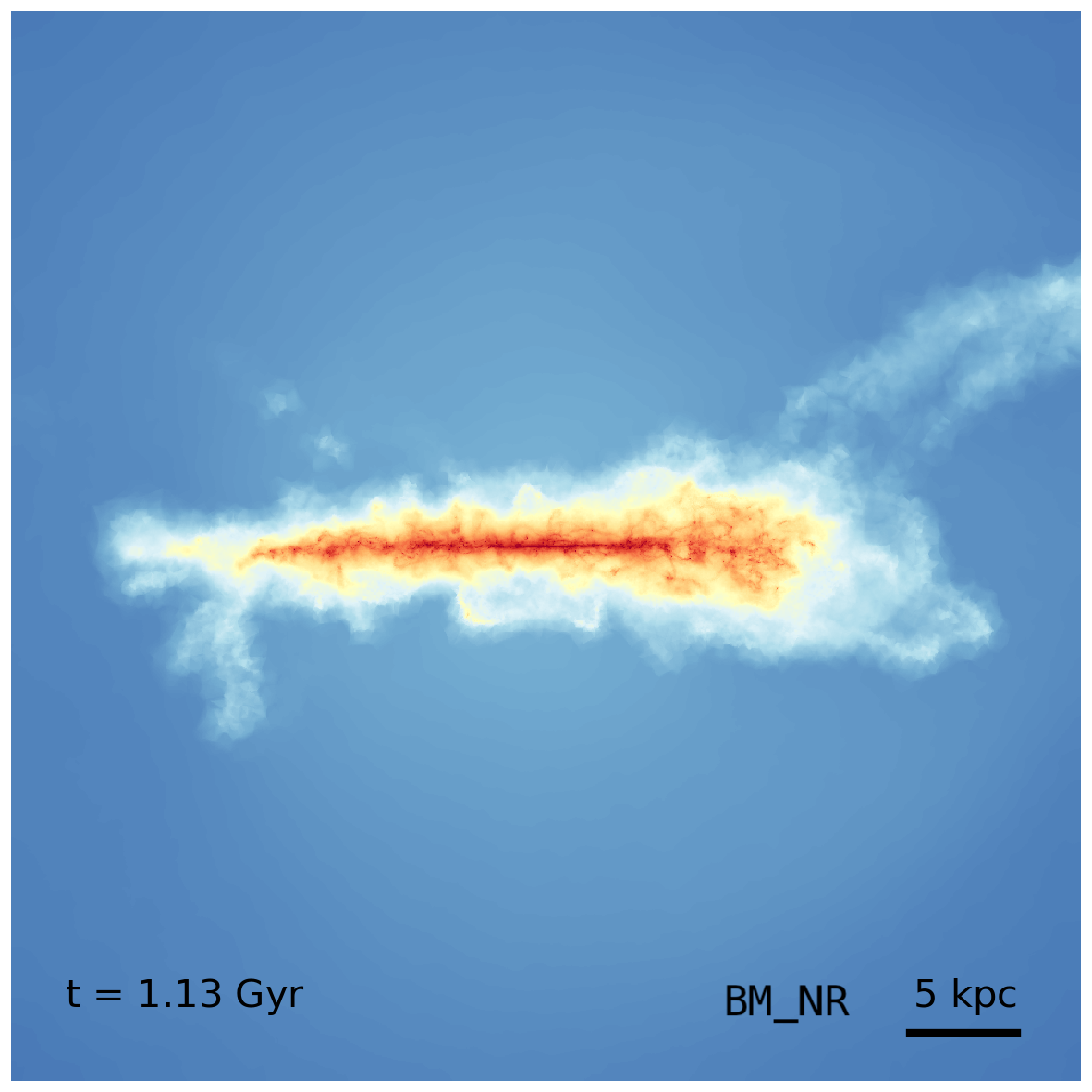}}

\caption{\textit{Top panels}: Gas column density in face-on (first row) and edge-on (second row) projections of the {\sc BM\_r} simulation computed using the \smuggle\ model at the times indicated in each panel. The horizontal dashed lines represent three heights over the disc (2, 5 and 10 kpc). Each panel is 50 kpc across and in projection depth with a total number of 1024 × 1024 pixels that give a resolution of $\sim 50$ pc. Redder colors correspond to higher densities, as indicated in the colorbar. From the face-on projections we can appreciate the presence of a complex ISM structure, with low density cavities produced by stellar feedback and high density filaments where the star formation takes place. The edge-on projections show the presence of the galactic corona around the disc and the generation of galactic-scale outflows. \textit{Bottom panels}: The same projections are made face-on (third row) and edge-on (forth row) at $t=1.13$ Gyr for the {\sc nocor} (first column), {\sc nofeed\_r} (second column) and {\sc BM\_nr} (third column) simulations. The structure of the gaseous disc and the outflows is remarkably different comparing these simulations with {\sc BM\_r}.}
\label{face1}
\end{figure*}

\begin{figure*}
\centering
\subfloat{\includegraphics[width=0.33\textwidth]{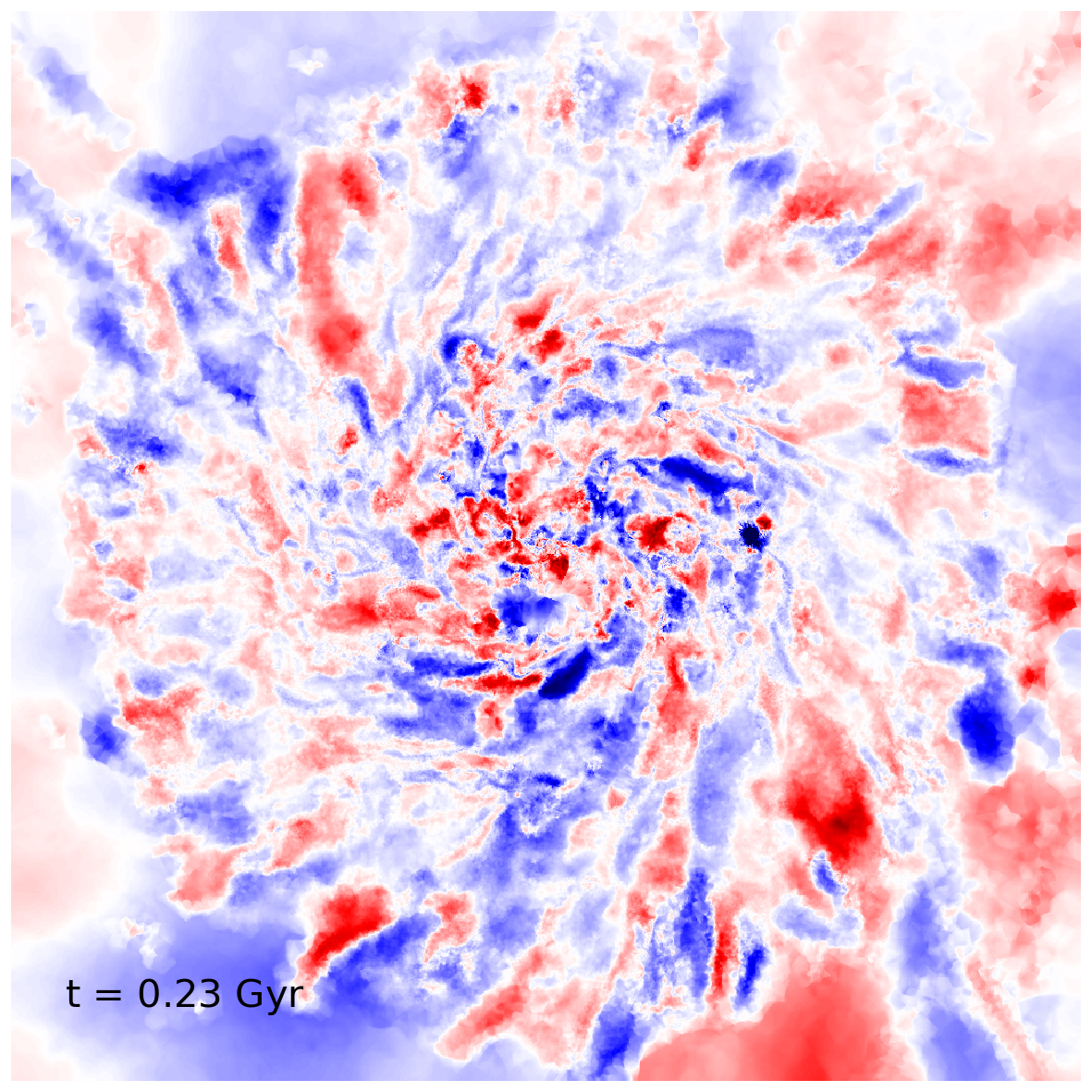}} 
\subfloat{\includegraphics[width=0.33\textwidth]{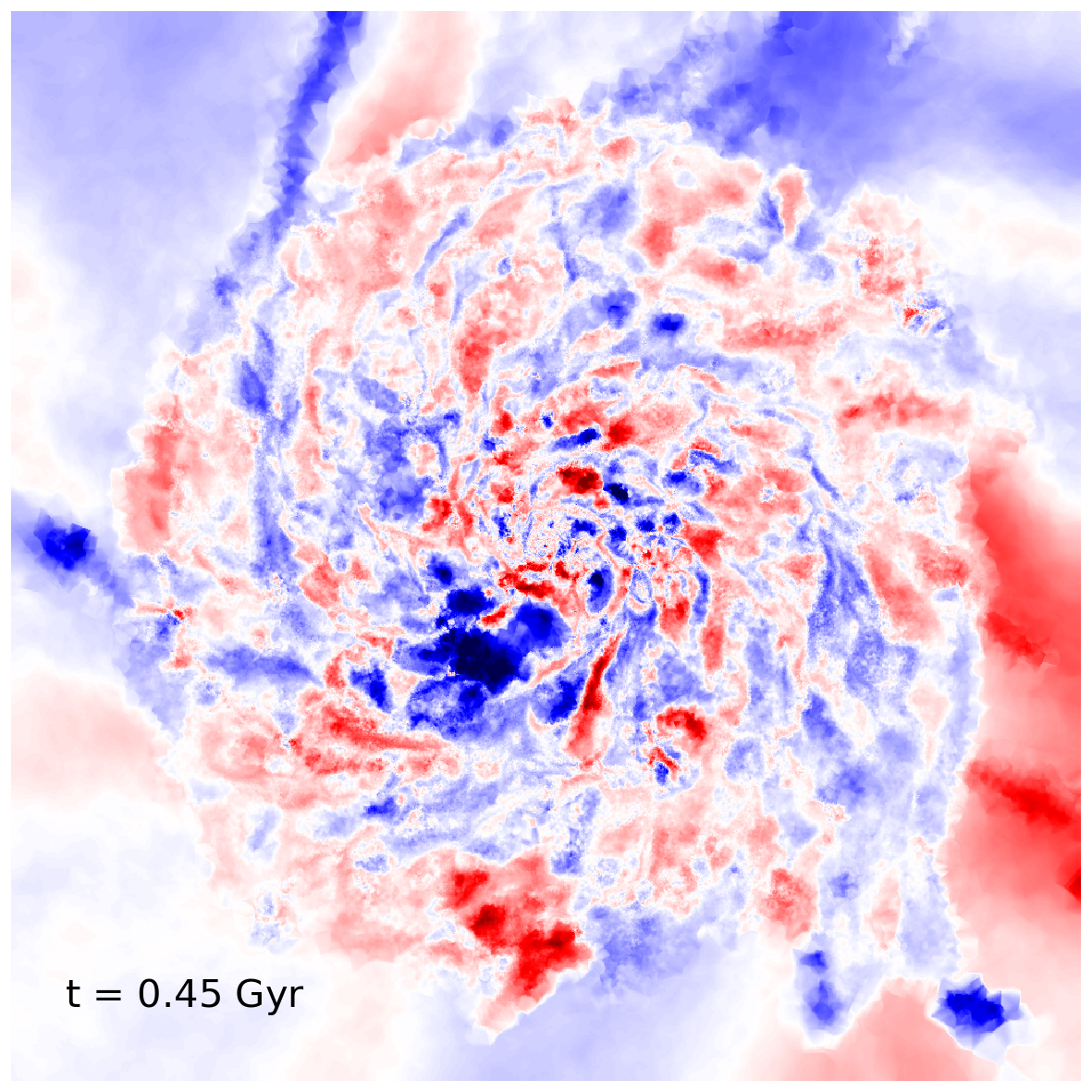}}
\subfloat{\includegraphics[width=0.33\textwidth]{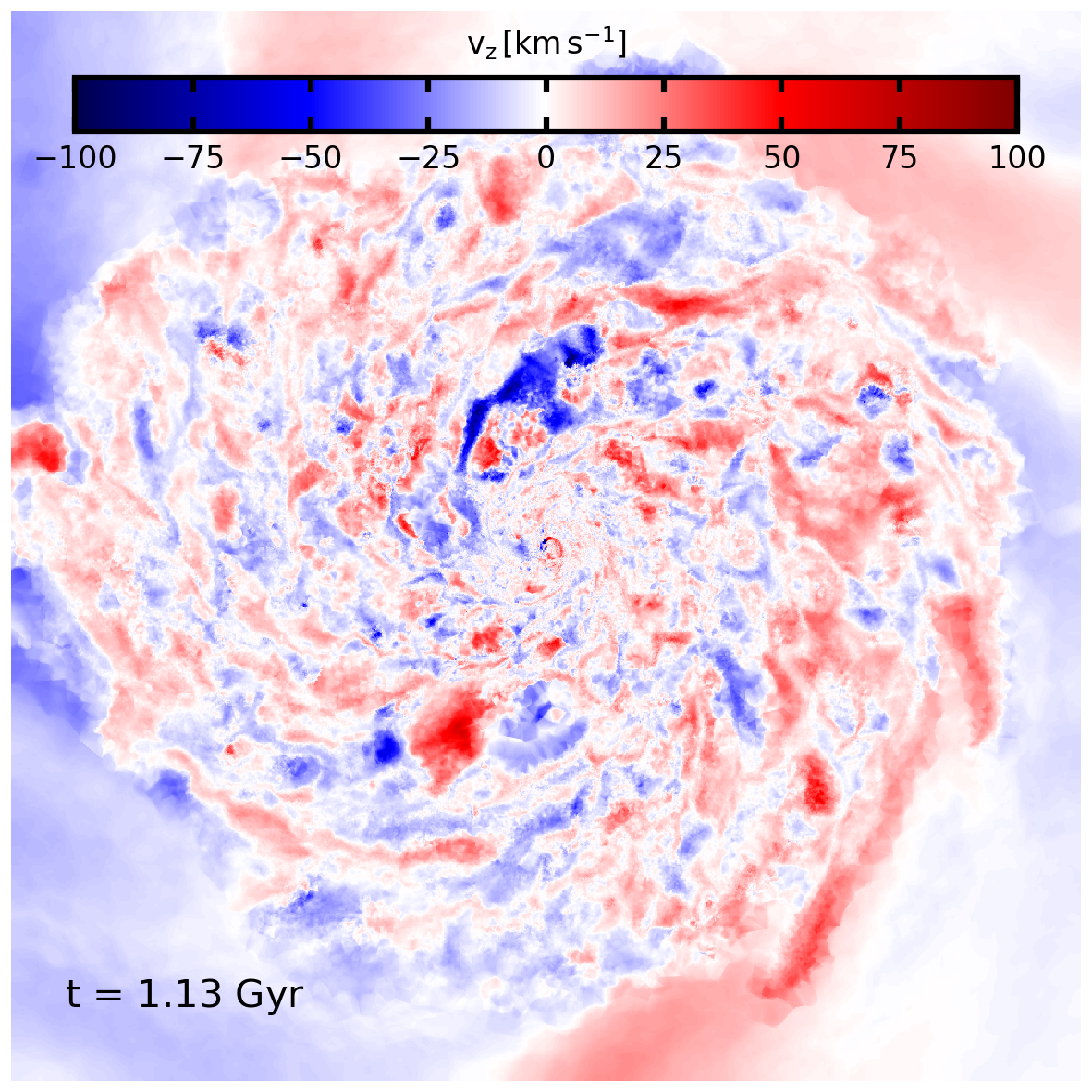}} \\[-3ex] 
\subfloat{\includegraphics[width=0.33\textwidth]{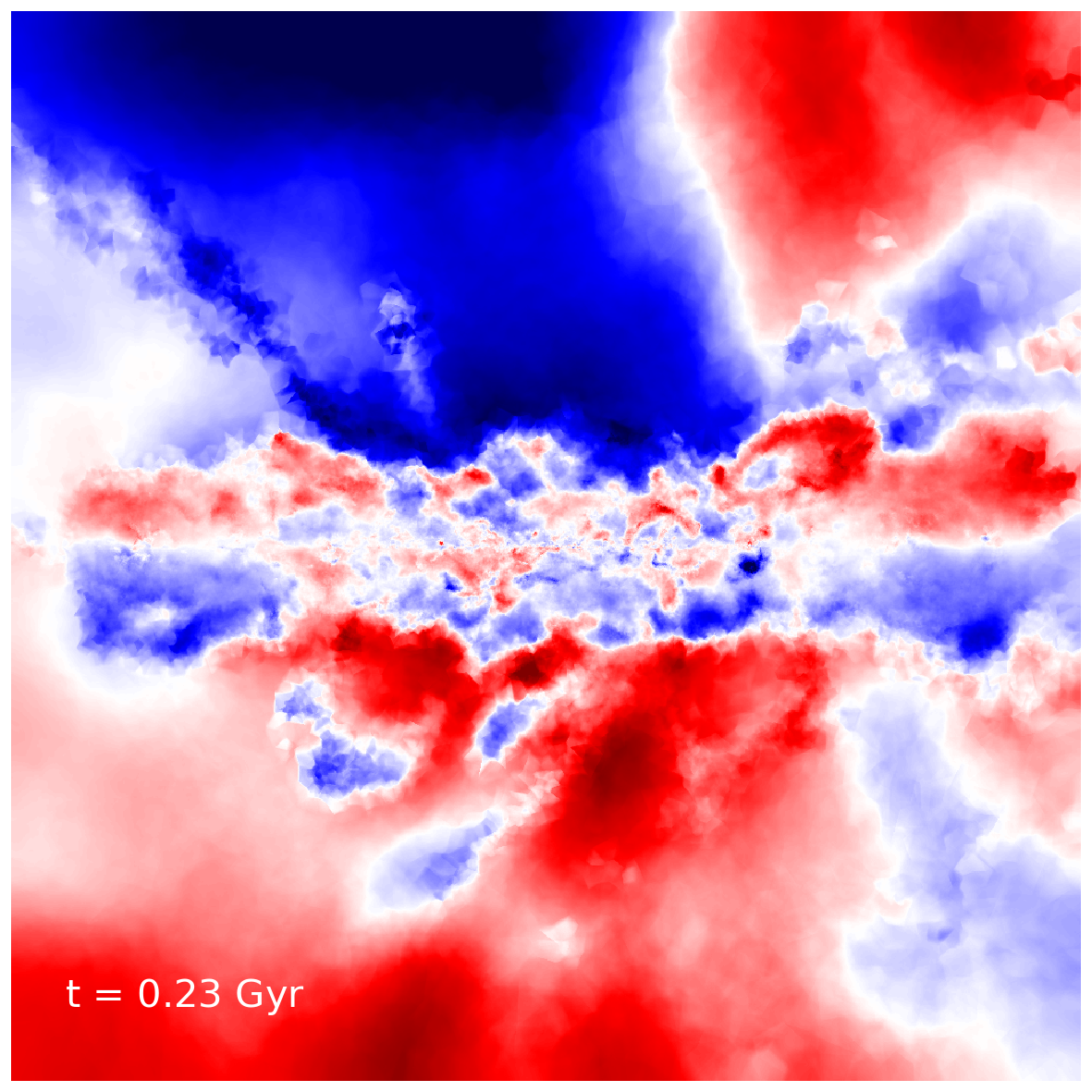}}  
\subfloat{\includegraphics[width=0.33\textwidth]{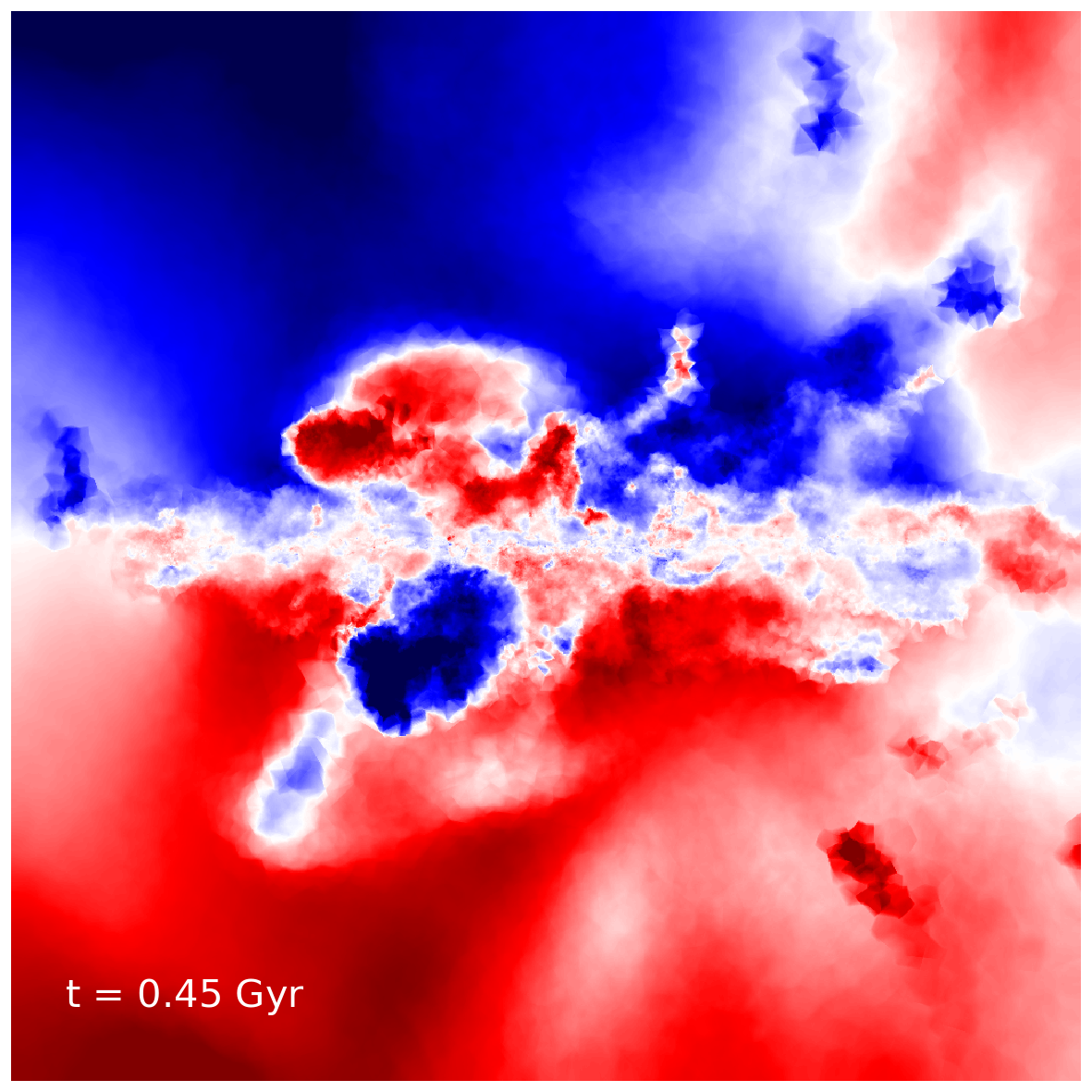}}
\subfloat{\includegraphics[width=0.33\textwidth]{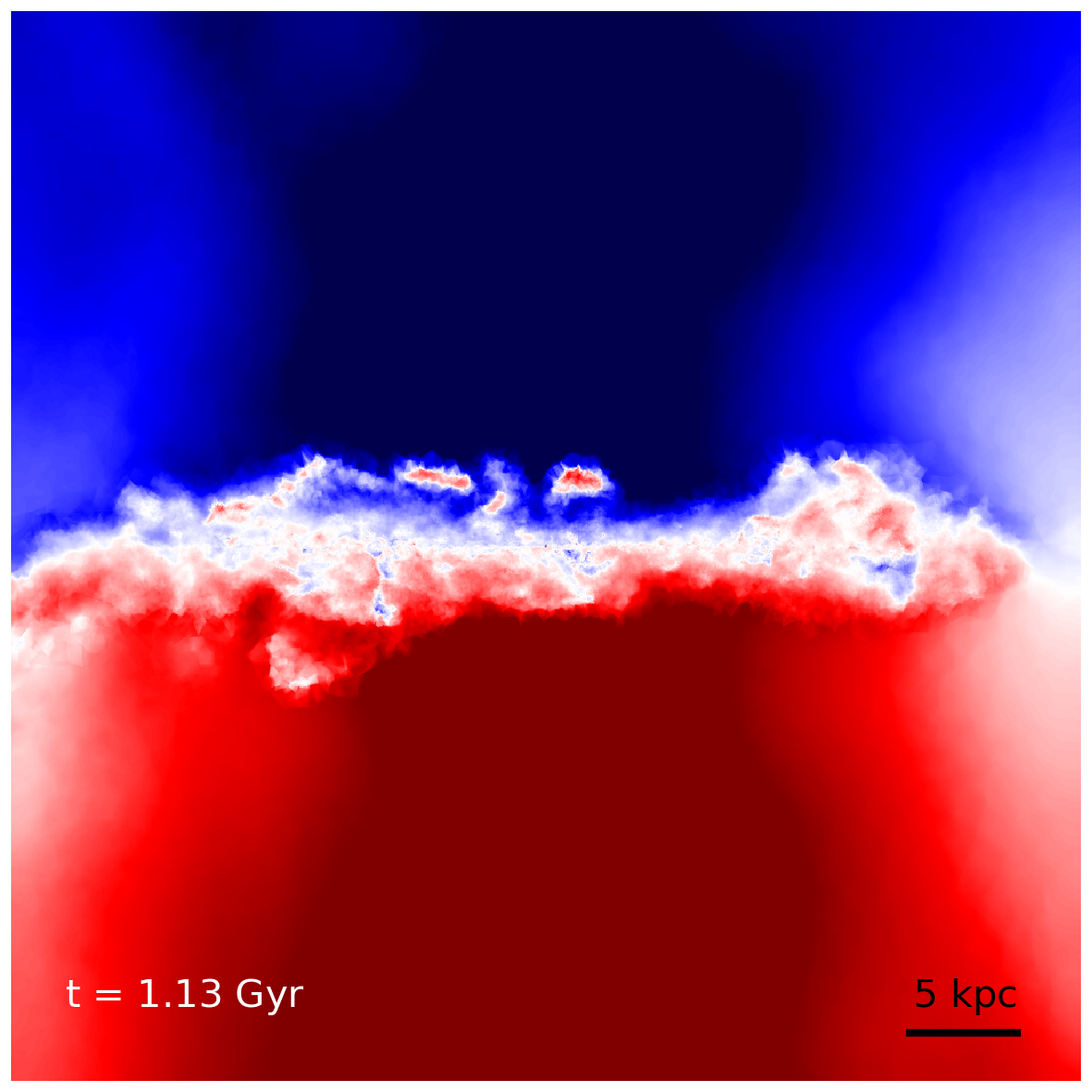}}

\caption{Density-weighted maps of the gas velocity perpendicular to the disc plane in face-on (top panels) and edge-on (bottom panels) projections of the {\sc BM\_r} simulation computed using the \smuggle\ model at the times indicated in each panel. Each panel is 50 kpc across and in projection depth with a total number of 1024 × 1024 pixels that give a resolution of $\sim 50$ pc.  We can visually appreciate the kinematic structure of the gas in the galaxy. The coronal gas is accreted onto the galaxy and outflows of gas are ejected from the disc, this generates a gas circulation between the disc and the corona.}
\label{edgevel}
\end{figure*}

\begin{figure*}
\centering
\subfloat{\includegraphics[width=0.33\textwidth]{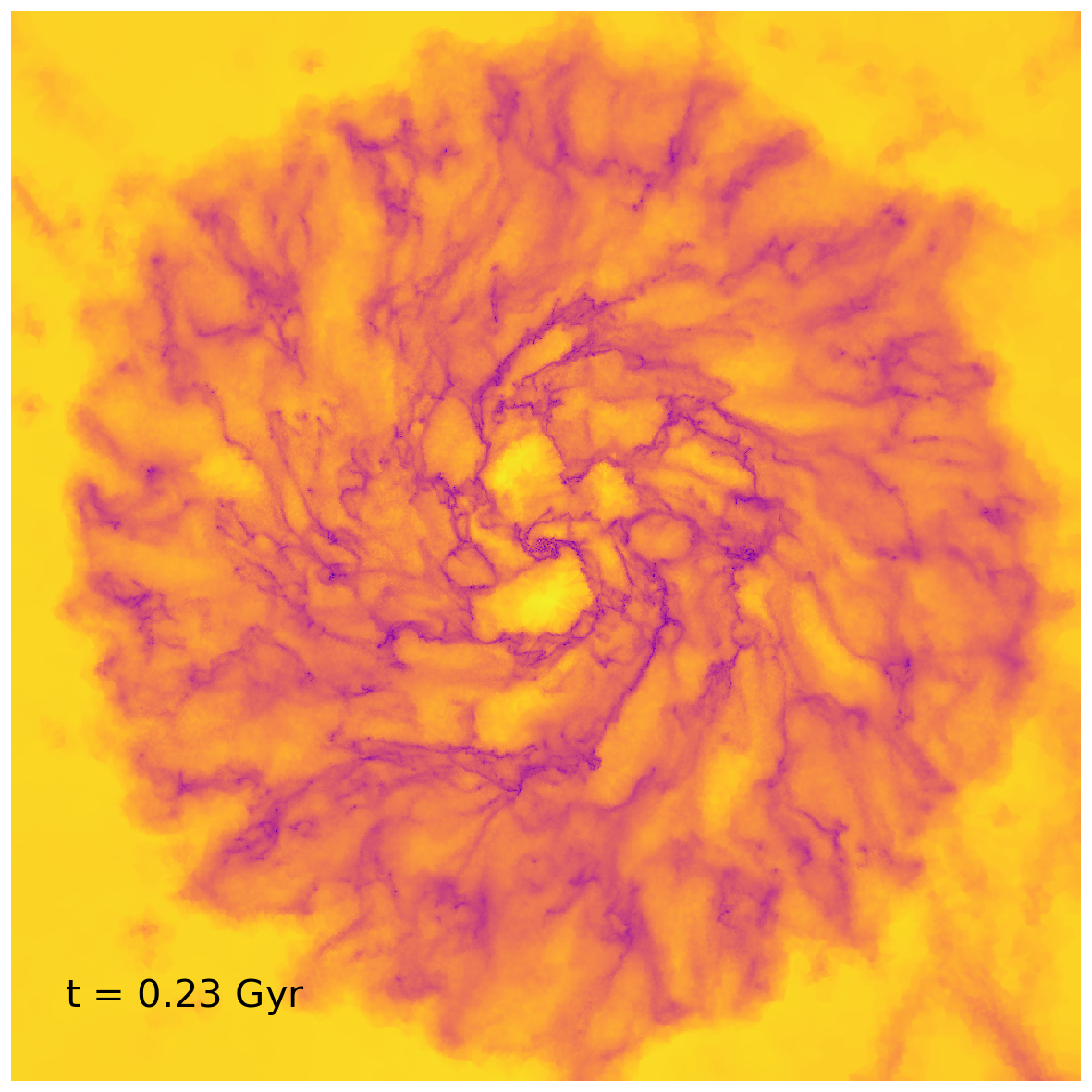}} 
\subfloat{\includegraphics[width=0.33\textwidth]{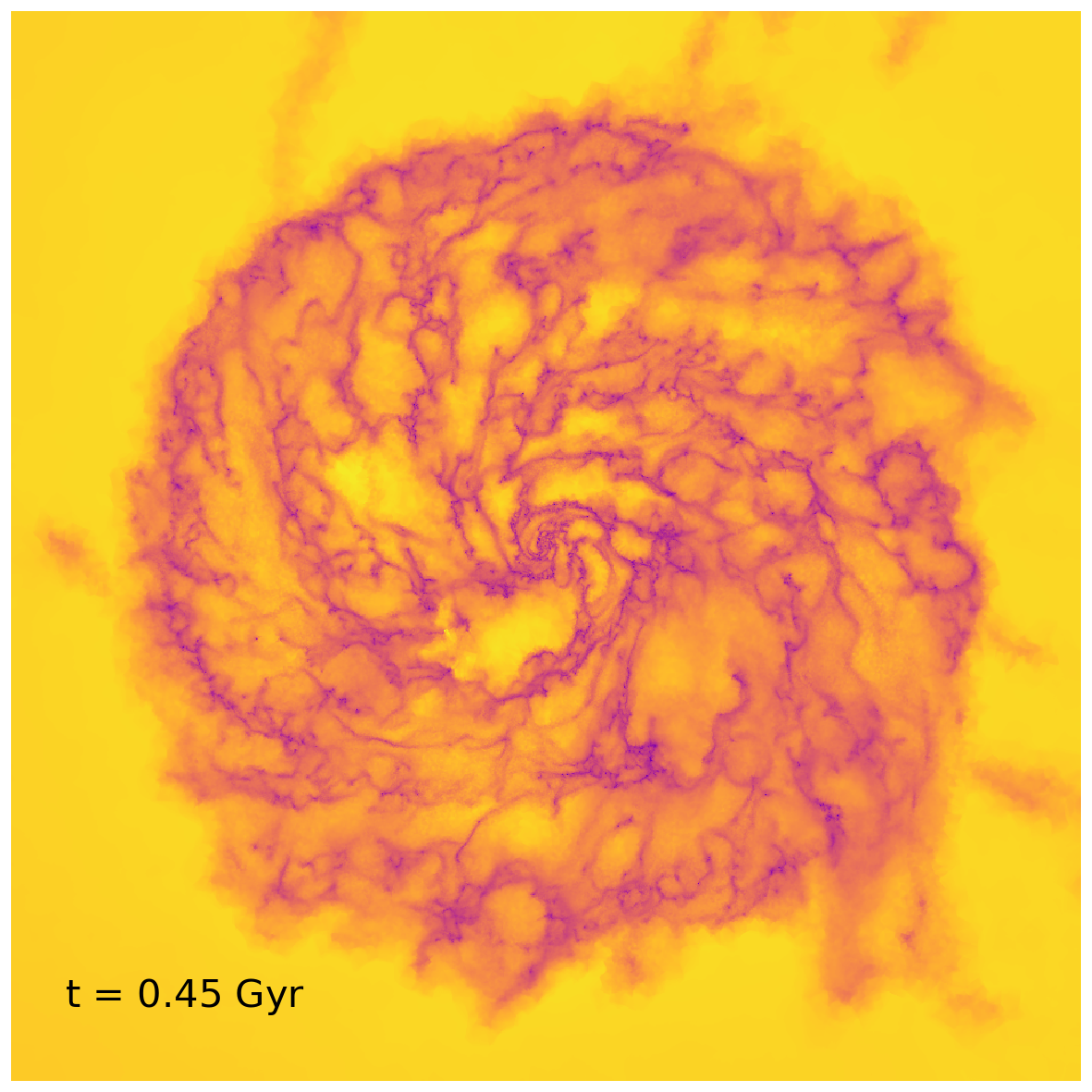}}
\subfloat{\includegraphics[width=0.33\textwidth]{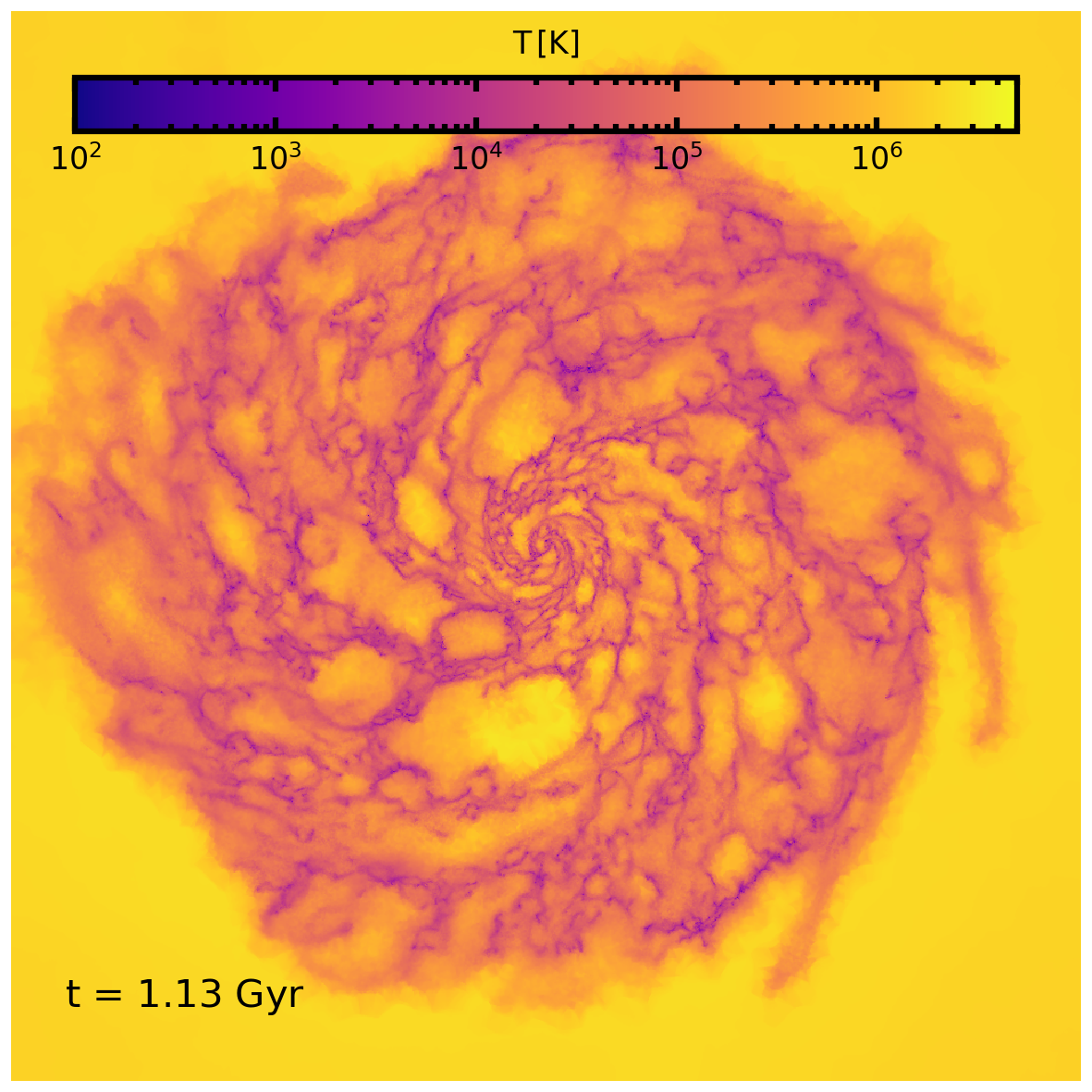}} \\[-3ex] 
\subfloat{\includegraphics[width=0.33\textwidth]{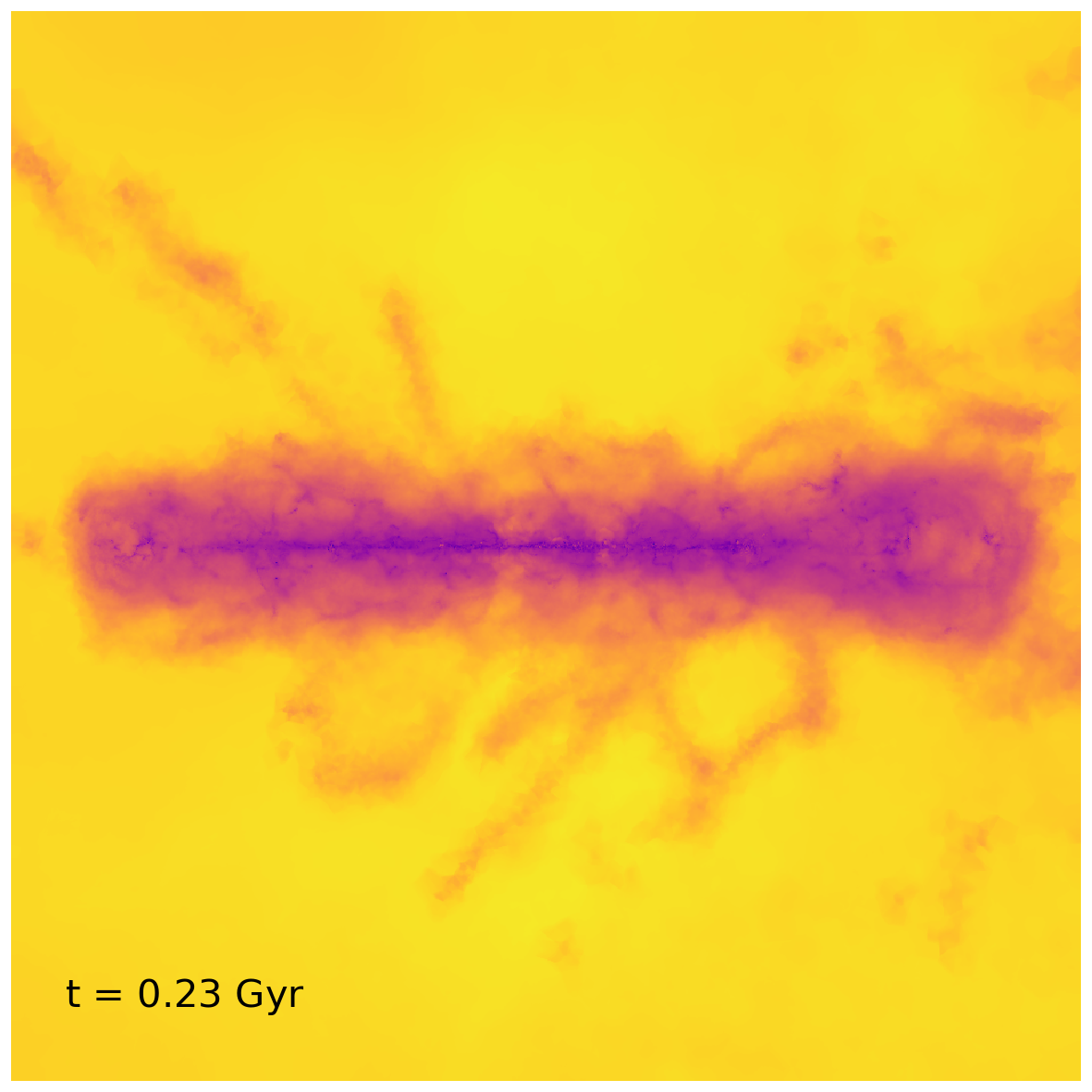}}  
\subfloat{\includegraphics[width=0.33\textwidth]{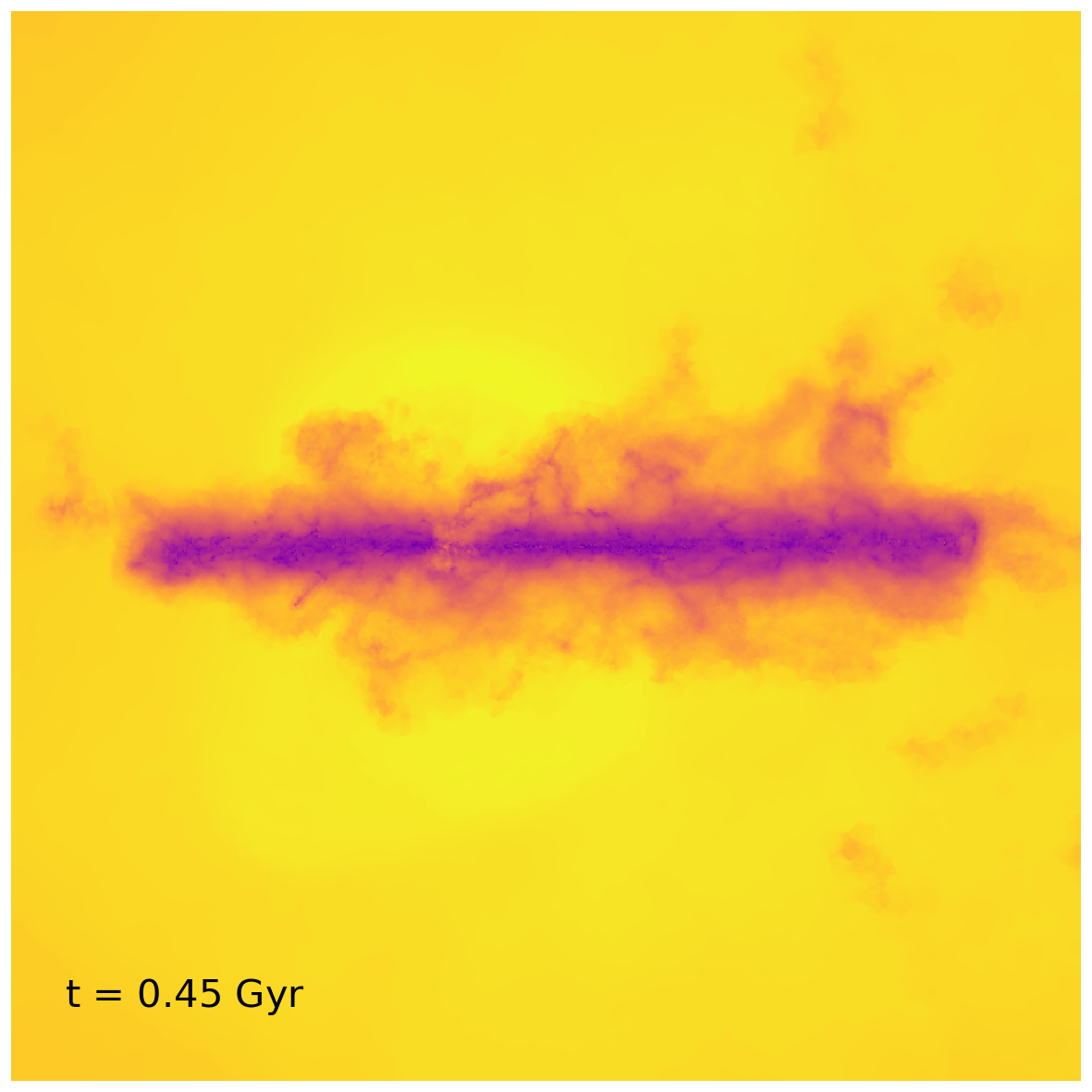}}
\subfloat{\includegraphics[width=0.33\textwidth]{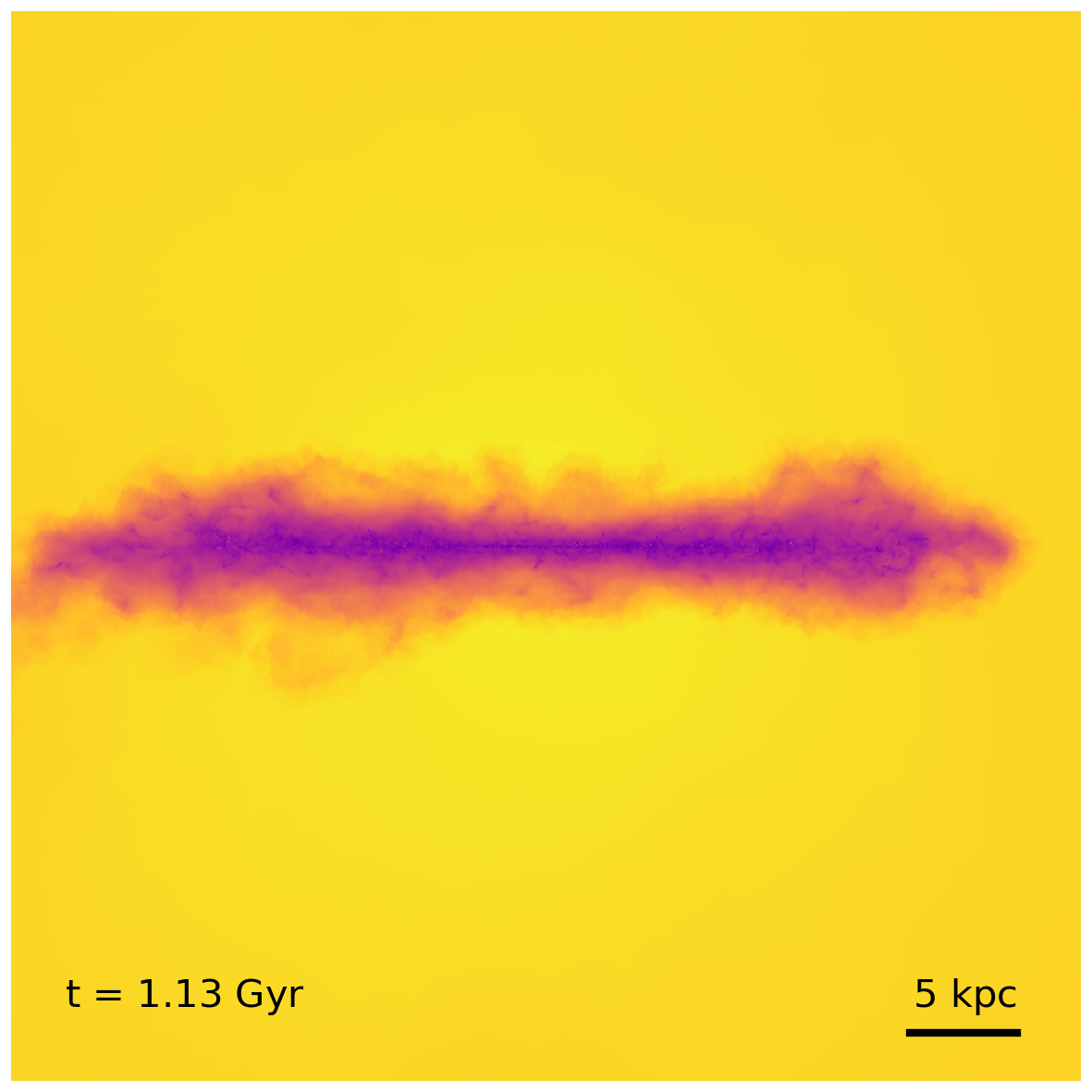}}

\caption{Density-weighted temperature in face-on (top panels) and edge-on (bottom panels) projections of the Milky-Way like galaxy for the {\sc BM\_r} simulation computed with the \smuggle\ model at the times indicated in each panel. Each panel is 50 kpc across and in projection depth with the presence of a total number of 1024 × 1024 pixels that give a resolution of $\sim 50$ pc. We can visually appreciate the mixing between the material ejected from the disc and the hot galactic corona, with the formation of an intermediate temperature gas phase at the disc-corona interface.}
\label{temp_proj}
\end{figure*}

For our simulations, we analysed two distinct configurations: a galaxy with a static corona and a galaxy with a rotating one. In the case of a static corona, the temperature profile is derived assuming that the coronal gas is in hydrostatic equilibrium in the total potential of the galaxy. Knowing the corona density profile $\rho_{\text{cor}}(r)$ and the total cumulative mass profile $M(r)$, the temperature profile is given by

\begin{equation}
    T(r) = \frac{\mu m_p}{k_B}\frac{1}{\rho_{\text{cor}}(r)}\int_{r}^{\infty} \rho_{\text{cor}}(r) \frac{G M(r)}{r^2} \ \text{d}r.
    \end{equation}
We also assign a uniform metallicity, set to $0.1 \ Z_{\odot}$, to the coronal gas \citep[e.g.][]{tosi1988, bogdan2017}.
In the case of a rotating corona, the net velocity is computed as a fraction of the circular velocity
\begin{equation}
    v_{\phi}=\alpha \biggl[\frac{\partial \Phi}{\partial R} R\biggr]^{1/2},
\end{equation}
where we set $\alpha=0.4$ in our runs, which ensures that $v_{\phi} \approx 90$ km s$^{-1}$ near the disc ($z\approx 2$ kpc). In fact, the galactic corona is expected to rotate with a lag of $\approx 80-120$ km s$^{-1}$ with respect to the disc of the galaxy \citep[e.g.][]{marinacci2011, afruni2022}. Values $\alpha > 0.4$ can cause significant deviations from sphericity, which, for simplicity is assumed in setting all the profiles of the coronal properties. A more accurate treatment of the rotation structure will be addressed in future works. The temperature profile can be computed assuming hydrostatic equilibrium in an effective potential
\begin{equation}
 \Phi_{\text{eff}} = \Phi(R, z) - \int_{\infty}^{R} \frac{v_{\phi}(R')}{R'} \ \text{d} R'.
\end{equation}
In both the static and the rotating configurations the temperature is higher in the centre and it decreases with radius (see purple dashed line in Figure \ref{hqbm}): a temperature gradient has been observed for instance in \citet{dai2020}, this is very similar to the temperature that we obtain in our ICs, with an average temperature $\sim 10^6$ K that depends on the total mass of the galaxy.

\begin{table}
	\centering
	\caption{Summary of the model variation of the Milky Way-like galaxy parameters in this paper. Our fiducial simulation is {\sc{BM\_r}}, which includes a rotating corona. The other simulations are used as comparison with the main one.}\label{model_var}
	\begin{tabular}[t]{cccc}
	
		\toprule
		\textbf{Simulation name} & $M_{\text{cor}}$ [M$_{\odot}$] & Rotation & Feedback\\
		\bottomrule
	    \rule{0pt}{3ex}  
		\sc{BM\_nr} & $3\times 10^{10}$& no & yes \\
		\sc{BM\_r} & $3\times 10^{10}$& yes & yes \\
		\sc{nocor} & negligible &  - & yes 
		\\
		\sc{nofeed\_{nr}} & $3\times 10^{10}$ & no & no \\
		\sc{nofeed\_{r}} & $3\times 10^{10}$ & yes & no \\
		 
		\bottomrule
	\end{tabular}
\end{table}%

\subsection{Simulations setup}
We simulated a series of isolated non-cosmological Milky Way-like galaxies, with components as described in the previous Sections. The parameters of the galaxies simulated in this work are listed in Table \ref{tabella_param}. These parameters are chosen in order to represent a galaxy similar to the Milky Way. The galactic corona has a mass inside the virial radius of $M_{\text{cor}}(r<r_{\text{vir}})=3\times 10^{10}$ M$_{\odot}$, that is a mass similar to those derived from observations \citep[e.g.][]{kaaret2020, martynenko2022} and is in agreement with the observational constraints on the number density (see Figure \ref{hqbm}) of the galactic corona in the Milky Way. The galaxy is inserted inside a box with a side length of $860$ kpc. We have run 4 main simulations (listed in Table \ref{model_var}) changing the characteristics of the coronal gas or the feedback processes. These are:
\begin{itemize}
    \item {\sc BM\_nr}: using a $\beta$-profile with no net rotation;
    \item {\sc BM\_r}: using a $\beta$-profile with rotation;
    \item {\sc nocor}: ICs with a negligible galactic corona;
    \item {\sc nofeed\_nr/nofeed\_r}: simulations with no stellar feedback.
\end{itemize}
 The {\sc BM\_r} simulation has been chosen as the fiducial simulation for this paper, it includes a rotating corona with a beta model density profile. The results of this simulation will be compared with a static corona case ({\sc BM\_nr}), a configuration without the galactic corona ({\sc nocor}) and two configurations in the absence of stellar feedback ({\sc nofeed\_nr/nofeed\_r}) with the same ICs of {\sc BM\_nr/BM\_r}.
We put $N_{\text{disc}}=3.2\times 10^6$ particles in the stellar disc, $N_{\text{bulge}}=8\times 10^5$ in the bulge and $N_{\text{gas}}\simeq 4.9\times 10^5$ gaseous particles, with $N_{\text{gas,d}}=3.2\times 10^5$ gaseous particles in the disc and $N_{\text{gas,cor}}=1.7\times 10^5$ in the corona. This results in a mass resolution $m_{\text{gas}}=1.1\times10^4$ M$_{\odot}$ and a minimum gravitational softening length $\epsilon_{\text{gas}} = 10$ pc for the gas. The mass of the stellar disc and bulge particles is respectively $m_{\text{disc}}=1.5\times10^4$ M$_{\odot}$ and $m_{\text{bulge}}=2\times10^4$ M$_{\odot}$, and the gravitational softening length is $\epsilon_{\star} = 21.4$ pc. 

First, we have run adiabatic simulations using \arepo, i.e. without the presence
of the dissipation processes implemented in the \smuggle\ model, with the different IC setups, to ensure that the different components of the galaxy are in equilibrium. This simulations confirmed the dynamical stability of the simulated galaxy
model, with a configuration that remains almost unchanged for a time span of 2
Gyr. Then the ICs are evolved using \arepo\ with the \smuggle\ model (Section \ref{simcode}) for 2 Gyr, this is a sufficient time to see the effect of the corona presence on the SFR of the galaxy, given that the depletion time for a Milky Way-like galaxy is
\begin{equation}\label{deptime}
\tau_{\text{dep}}=\frac{M_{\text{g}}}{\text{SFR}}\sim \frac{10^9 \ M_{\odot}}{1 \ M_{\odot} \ \text{yr}^{-1}} \sim 1 \ \text{Gyr},
\end{equation}
where $M_{\text{g}}$ is the mass of gas in the disc and SFR is the star formation rate, $\tau_{\text{dep}}$ has similar values also in our setup.

\section{ISM and coronal gas structure}\label{res}

\subsection{Scheme of the gas circulation}

In Figure \ref{galaxy_scheme} we show a schematic view of the processes that contribute to the gas circulation, i.e. the balance between outflows and inflows (from and into the galactic disc), in a star-forming galaxy like the Milky Way. The gas that is lifted from the disc (outflowing) is ejected thanks to stellar feedback ($\sim \dot{M}_{\text{out}}$), mostly coming from SN explosions. The gas accreted onto the galactic disc (inflowing) can be schematically divided in two main components: (i) the galactic fountain gas that, failing to reach the galaxy escape velocity, re-accrete onto the disc ($\dot{M}_{\text{in, fount}}$, which is $\sim \dot{M}_{\text{out}}$), (ii) the gas accreted from the galactic corona that comes directly from the radiative cooling of the coronal gas ($\dot{M}_{\text{in, cor}}$) or from the interaction between the galactic fountains and the corona ($\dot{M}_{\text{in, int}}$). The latter interaction may act as a positive feedback for star formation, as mentioned in Section \ref{introduction}. The coexistence of these different phenomena is described in more detail in the following Sections.

\begin{figure*}
	\includegraphics[width=\textwidth]{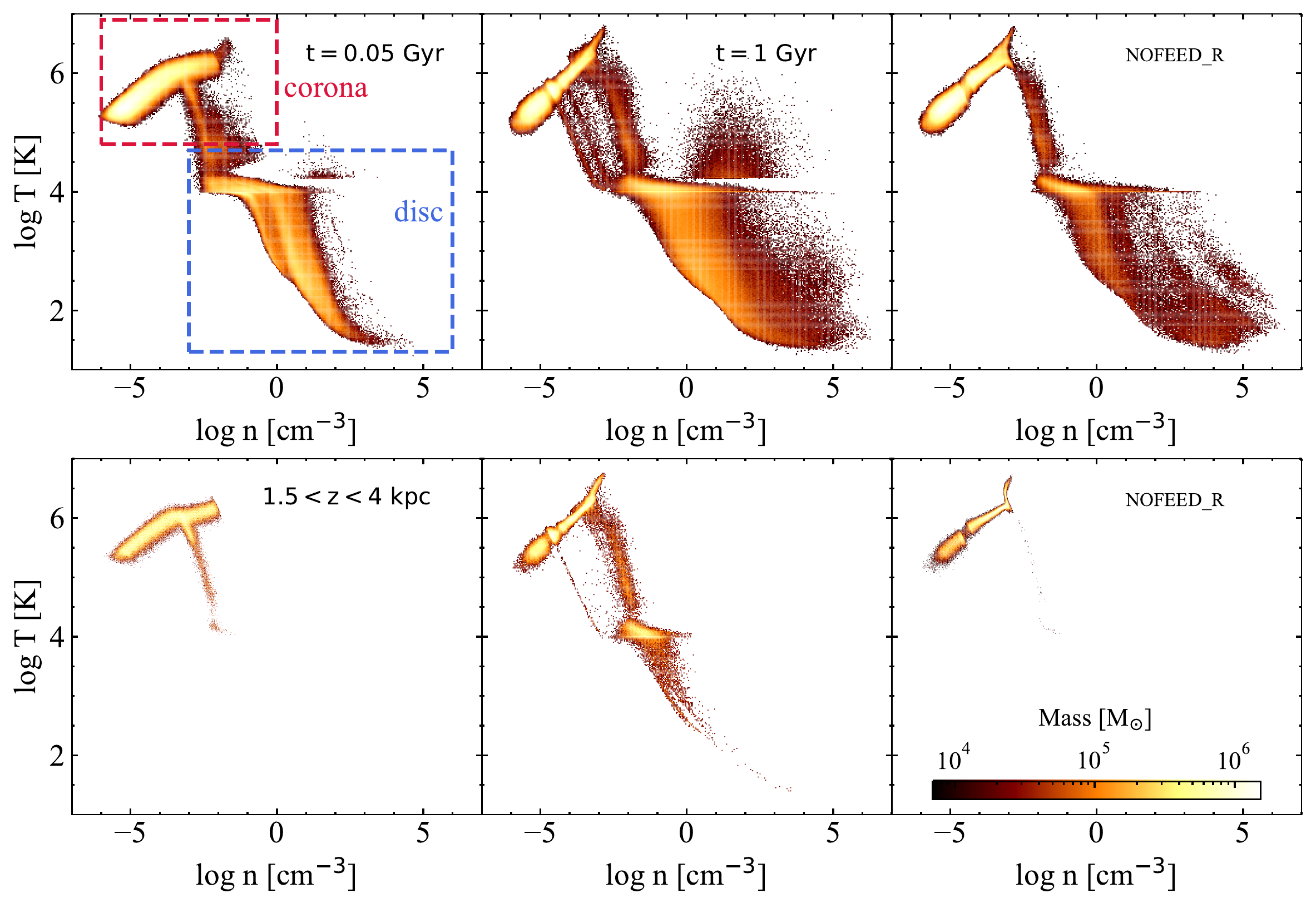}
    \caption{Number density-temperature diagrams for all the gas cells in the simulation box (top panels) and for the gas cells at the disc-corona interface ($1.5<z<4$ kpc, bottom panels) of the {\sc BM\_r} simulation at $t=0.05$ Gyr (first column) and $t=1$ Gyr (second column) and of the {\sc nofeed\_r} simulation at $t=1$ Gyr (third column). The lighter colours indicate higher gas masses. The blue and the red dashed rectangles specify the gaseous disc and the coronal gas respectively. The diagram highlights the complex structure of the ISM and of the coronal gas. The gas in the disc has a wide range of temperatures and densities thanks to the cooling processes and to the stellar feedback. We can see the structure of the corona, that cools starting from the more dense and hot gas located at the centre of the galaxy. A different structure is present in the gas at the interface between {\sc BM\_r} and {\sc nofeed\_r}.}
    \label{phase_diagram}
\end{figure*}

\subsection{Gas projection maps}
We start by analysing projection maps of the gas in the simulated galaxies. In Figure \ref{face1} we show the face-on and edge-on surface density maps of our fiducial simulation {\sc BM\_r} at different times and a comparison with other simulations. From the face-on projections of {\sc BM\_r} (first row in Figure \ref{face1}) we can appreciate the gaseous disc structure complexity of the simulated galaxy: as the simulation starts, the first stars begin to form, generating radiative feedback and stellar winds that are followed by the SN explosions, which cause the formation of low-density cavities (surface densities $<1 M_{\odot} \ \text{pc}^{-2}$). These cavities, filled with hot gas, can extend over the plane of the disc and are responsible for the production of outflows. Along the spiral arms of the galaxy, at the edges of the generated cavities, high-density gas filaments, whose presence is enhanced by stellar feedback, are present. In such regions, most of the star formation takes place, since below the density threshold $\rho_{\text{th}}$ the formation of a stellar particle is not possible (see Section \ref{smuggle}).

In Figure \ref{face1} we have also compared the disc structure of {\sc BM\_r} with the other simulations performed at $t=1.13$ Gyr (third row). The first panel shows the {\sc nocor} simulation, where the disc is more extended owing to the lack of accretion of low-angular momentum gas from the corona. Furthermore, the overall gas surface density is visibly lower, especially in the gas filaments: the gas consumed by star formation is not replaced and the average gas density can only decrease over time. The {\sc nofeed\_r} simulation is shown in the second panel. Here, the disc structure is unrealistic with almost no gas filaments and with small clumps of high-density gas. These clumps are formed from initial density fluctuations thanks to gravitational attraction, they grow over time and are not torn apart from SN explosions. Also, they tend to migrate towards the centre of the disc over time as a consequence of dynamical friction. This phenomenon causes a higher concentration of gas at the centre of the disc, where a substantial amount of stars are then able to be formed. A few large low density cavities are still formed owing to gas clustering and cooling. Looking at the {\sc BM\_nr} simulation (third panel), we note that the gaseous disc scale length is smaller (halved approximately) then in {\sc BM\_r}. The disc shrinks in size over time in the case of a static corona, whereas it remains extended if the rotation is included. Although this might be caused by a redistribution of the angular momentum between the disc and the corona caused by the gas circulation mediated by the galactic fountains, this speculation requires further investigations and it is out of the purposes of this work. Overall, the disc structures found in {\sc BM\_r} are also present in this simulation. 

The second row of Figure \ref{face1} shows the corresponding edge-on projections of the same configurations shown in the top panels. In the first panel we have highlighted three heights over the plane of the disc (2, 5 and 10 kpc) as comparison. As we will discuss in Sections \ref{outin}, \ref{pos_feed} and \ref{outt}, at these heights we have computed the outflow and inflow rates (inside the shaded area). We can observe the presence of a low density halo of gas around the galactic disc. In an initial phase that lasts about $0.1$ Gyr there is a substantial accretion of coronal gas onto the disc, as we will discuss in detail in Section \ref{starform}. This accretion episode causes an increase of the SFR at the beginning of the simulation, after this the corona inflates again and continues its evolution. Starting from $t=100$ Myr the galaxy develops strong outflows, ejected from the disc by stellar feedback. These outflows can reach relatively large distances from the galaxy mid-plane (over $10$ kpc), travelling through the galactic corona and then falling back onto the disc in a galactic fountain cycle \citep[e.g.][]{fraternali2006}. This is clear in the 0.23 Gyr image where we can see several galactic fountains ejected from the disc. This process is schematically depicted in Figure \ref{galaxy_scheme} and will be deeply analysed in Sections \ref{pos_feed} and \ref{outt}.

We have analysed the gaseous structure of the other simulations at $t=1.13$ Gyr in comparison with {\sc BM\_r} also for the edge-on projections (see forth row in Figure \ref{face1}). In {\sc nocor} (first panel) the disc is isolated and not surrounded by the corona, the intensity of the gaseous outflows is lower than in {\sc BM\_r}. The {\sc nofeed\_r} simulation (second panel) features a disc immersed in a galactic corona, but in the absence of galactic outflows. From the {\sc BM\_nr} simulation (third panel) we note that the gaseous outflows are more frequent and reach higher heights with a static corona: with a rotating corona the gas surface density is lower due to the larger scale length of the gaseous disc induced by the angular momentum exchange. Moreover, the rotation of the gas creates a different gas accretion distribution on the disc. For these reasons, the star formation is lower and consequently also the intensity of gas ejection. A more detailed outflow/inflow analysis is carried out in Section \ref{outin}.

Figure \ref{edgevel} shows the face-on and edge-on projections of the density-weighted gas velocity perpendicular to the disc plane of the {\sc BM\_r} simulation. The bluer colors in the region above the galactic disc mid-plane represent gas that is falling onto the disc, the redder colors gas that is ejected from the disc, and vice versa for the regions below the disc. The face-on projections (top panels) highlight the gas that is ejected at high velocity from SN explosions: the location of this ejected gas roughly corresponds to the low-density cavities in Figure \ref{face1}. In the edge-on projections (bottom panels) is particularly noticeable the accretion of coronal gas that is strongest in the central regions, where its density is the highest. In addition, we see the complex kinematic structure of the gaseous outflows visible in Figure \ref{face1}: these outflows can have a velocity in excess of $100$ km s$^{-1}$. The galactic fountain kinematics is clearly visible in the projections at 0.23 and 0.45 Gyr. There is the establishment of a circulation of gas between the disc and the galactic corona: the gas is ejected at high velocities from the disc plane (with a vertical velocity directed out of the disc) and after a certain time, that depends on the initial velocity and on the angle of ejection, it falls back into another spot (with a vertical velocity directed to the disc plane).

In Figure \ref{temp_proj} we show the density-weighted projected temperature of the gas for the {\sc BM\_r} simulation. In the face-on projections (top panels) we note the presence of a multi-phase medium in the disc, where the regions with the higher densities correspond to the colder gas ($T<10^2$ K), this very cold gas can therefore collapse and form stars. The bubbles carved by SN feedback are filled with high temperature ($T\sim 10^6$ K) gas. The temperature distribution of the gas ejected from the disc is visible in the edge-on projections (bottom panels). At 1 kpc above and below the disc the gas is hotter than in the disc filaments, reaching $10^4<T<10^5$ K. Outside the disc of the galaxy the temperature is dominated by the coronal gas ($T>5\times10^5$ K). The same features are present in the temperature projections of the other simulations. The temperature of the aforementioned gaseous outflows is in general at an intermediate level ($10^4<T<10^5$ K), this might be caused by the mixing with the gas from the corona, that is at higher temperatures ($T\sim 10^6$ K), or it is possible that the majority of the gas is already ejected at these temperatures. As we will discuss in Section \ref{starform_circ} the gas mixing has a predominant role, we found that the intermediate temperature gas is formed due to the presence of the stellar feedback that mediate the interaction between the disc and the corona.

\subsection{Phase diagrams}
We now focus on the phase diagram of the gas in the galaxy, looking in particular to the {\sc BM\_r} simulation. The analysis of the structure of the gas that surrounds and lies inside the disc of the galaxy is critical to understand how star-forming galaxies evolve. In Figure \ref{phase_diagram} we show the gas number density-temperature phase diagram that includes all the gas cells in the simulation (top panels) or all the gas cells within a distance from the mid-plane $1.5<z<4$ kpc (bottom panels). The diagrams are represented at the beginning of the simulation ($t=0.05$ Gyr, first column) and at $t=1$ Gyr for {\sc BM\_r} (second column) and for {\sc nofeed\_r} (third column). The lighter colours represent higher gas masses, as indicated in the colorbar. The first noticeable thing is the presence of a complex ISM structure (blue dashed rectangle) that is already developed in the first Myr of the simulation, similarly to \citet{marinacci2019simulating}. The gas has a wide range of temperatures and densities: the majority ($\sim 85 \%$) of the gas in the disc has a temperature lower than $10^4$ K, with densities ranging from $10^{-3}$ cm$^{-3}$ to $10^{3}$ cm$^{-3}$. The strip of gas in Figure \ref{phase_diagram} at $1.7 \times 10^4$ K (i.e. log$T=4.25$ K) at densities between 1 and $10^{5}$ cm$^{-3}$ is gas photoionized from the massive and hot OB stars. Around these stars the gas temperature is fixed at this value. The creation of lower temperature gas is fundamental for the star formation since only gas over certain densities ($\sim 10^{-1} - 10^{-2}$ cm$^{-3}$) can cool, overcoming the heating from the UV background. This gas can reach densities over $10^{2}$ cm$^{-3}$ and is used to form new stars if the right conditions are satisfied (Section \ref{smuggle}). The cloud of gas particles with a temperature over $1.7 \times 10^4$ K and a density between $1$ and $10^{5}$ cm$^{-3}$ represents the hot ionized medium generated from the SN explosions.

The majority of the gas in the system (over $75 \%$) is located in the galactic corona, its gas can be identified in Figure \ref{phase_diagram} within the region marked by the red dashed rectangle (top left region of the diagram), in both the first and second panels. The coronal gas has a temperature between $10^5$ K and $10^7$ K and a low density, smaller than $10^{-1}$ cm$^{-3}$ and reaching $n < 10^{-5}$ cm$^{-3}$. In the first two panel from left to right we can appreciate how the temperature and density of the hot gaseous halo change over time: a `cooling bridge' forms between the corona and the gas in the disc at around $10^4$ K. This represents the coronal gas that is cooling and is fueling the galaxy with new gas for the star formation. This gas can cool more rapidly owing to the mixing with the galactic fountains. The first coronal gas that cools down is the one present in the central regions of the corona, that has the higher densities and temperatures.

The {\sc nofeed\_r} simulation (third panel) has a similar overall structure, but some differences are clearly evident. The hot gas generated by the SNe is absent due to the lack of stellar feedback. Furthermore, a noticeable difference is present in the structure of the `cooling bridge': the structures at lower densities and temperatures (present in {\sc BM\_r}) are totally absent. If we look at the lower panels, we can observe the density-temperature diagram for the gas at the disc-corona interface ($1.5<z<4$ kpc). In {\sc BM\_r} at $t=1$ Gyr (second panel) a multiphase medium, with both warm and cold gas, as well as hot gas, is present, instead, this is missing at the start of the simulation (first panel, $t=0.05$ Gyr) since the first SNe are not exploded yet. In {\sc nofeed\_r} (third panel) the cold and warm gas phases are completely absent. This gives us a first idea on the importance of stellar feedback, and therefore the production of galactic fountains, for the creation of a multiphase medium over the plane of the disc. The formation of warm and cold gas does not seem to be caused exclusively by the coronal gas cooling, otherwise these phases of gas would be somehow present in {\sc nofeed}. This aspect will be better explored in Section \ref{outt}.

\section{Star formation and galactic gas circulation}\label{starform_circ}

\subsection{Star formation history}\label{starform}

\begin{figure}
	\includegraphics[width=\columnwidth]{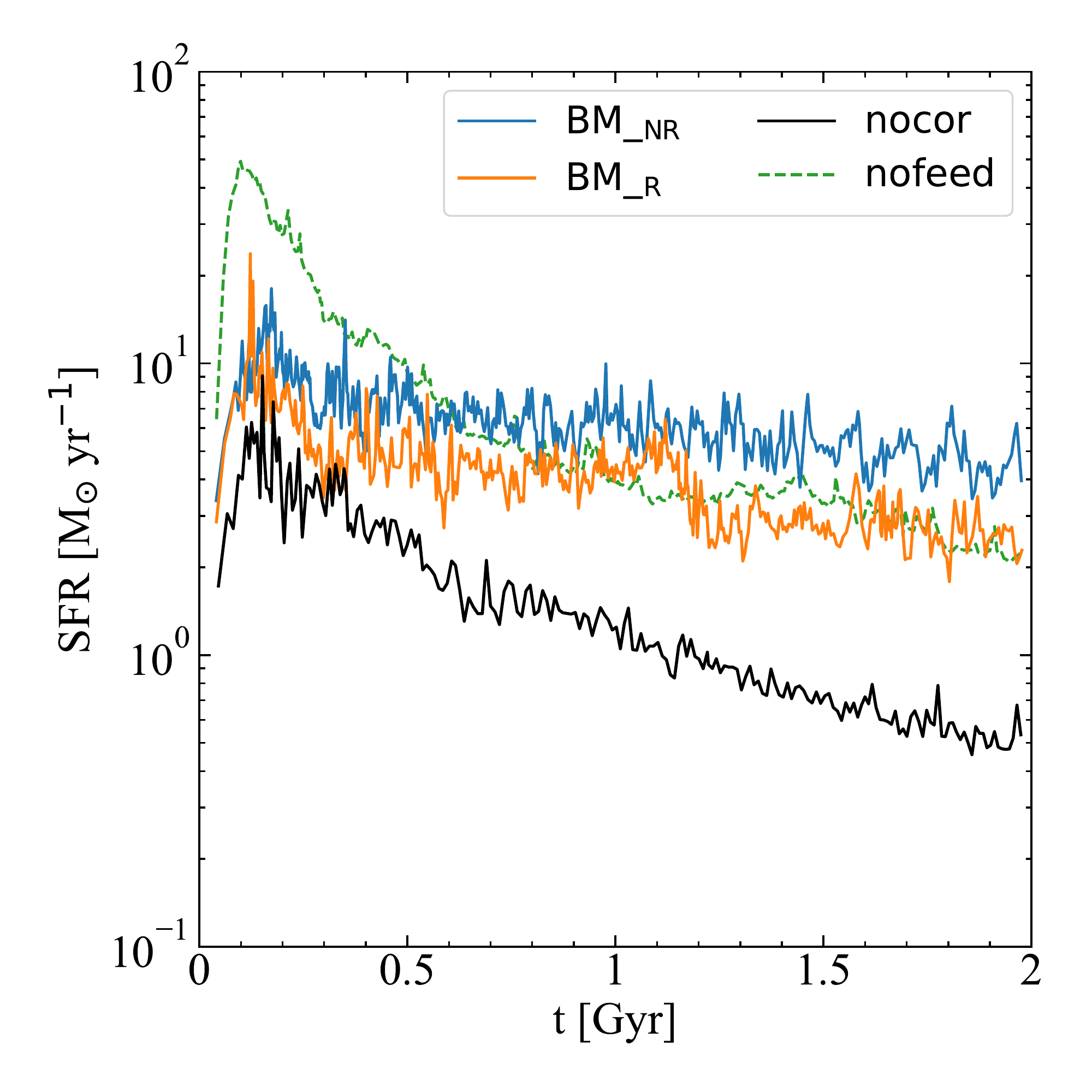}     
	\caption{Star formation rate (SFR) as a function of time for the four simulations performed: {\sc BM\_nr} (blue line), {\sc BM\_r} (orange line), {\sc nocor} (black line) and {\sc nofeed\_r} (green dashed line), see Table \ref{model_var}. In the {\sc nocor} simulation the SFR is decreasing over the time, reaching values of the SFR lower than {\sc BM\_nr}/{\sc BM\_r} at the end of the simulation. The simulations with the inclusion of the corona maintains an almost constant SFR up to 2 Gyr. {\sc nofeed\_r} features a prominent SFR peak within the first 0.5 Gyr.}
    \label{SFH}
\end{figure}

In Figure \ref{SFH} we show the evolution of the star formation rate (SFR) over time of the {\sc{BM\_nr}} (blue line) and {\sc{BM\_r}} (orange line) simulations, and compared it with the {\sc{nocor}} (black line) and {\sc{nofeed\_r}} (dashed green line) simulations. The evolution of the SFR in the first 0.2-0.3 Gyr has to be considered cautiously. We set the corona in hydrostatic equilibrium, but the activation of the cooling processes causes a significant deviation from it and causes the corona to lose pressure support in the central regions that rapidly accrete onto the disc. This initial accretion phase is due to the fact that the corona is already in place at the beginning of the simulations. A more realistic configuration would gradually form the hot corona over time following the processes mentioned in Section \ref{introduction}. After $\sim 0.2$ Gyr the SFR stabilizes, owing to the self-regulation between star formation and stellar feedback achieved by the \smuggle\ model. Also, the corona is more stable, but it slowly continues to accrete onto the disc owing to radiative cooling. 

We note the presence of a periodicity in the SFR, this `noisy' trend is linked to the self-regulation between the formation of stars and the stellar feedback that it causes. On the one hand, the periodic activation of the stellar feedback temporarily reduces the star formation injecting energy and momentum in the ISM, heating the gas and ejecting it from the disc. On the other hand, the periodic re-accretion of gas onto the disc can support the star formation with cold gas.

In the {\sc BM\_nr} simulation the SFR is $\approx 7-8 \ M_{\odot} \ \text{yr}^{-1}$ after the bump and it slightly decreases over time to $\approx 4-5 \ M_{\odot} \ \text{yr}^{-1}$ at 2 Gyr. The {\sc BM\_r} simulation has a similar trend, where the SFR stays at $ 2-3 \ M_{\odot} \ \text{yr}^{-1}$ after 2 Gyr (starting from $\approx 5-6 \ M_{\odot} \ \text{yr}^{-1}$ after the bump) until the end of the simulation. These SFR are in reasonable agreement with the SFR observed nowadays in the Milky Way \citep[e.g.][]{licquia2015} or in analogous galaxies \citep{mckelvie2019}. We note that the SFR is slightly higher when the coronal gas is not rotating: the rotation of the gas modifies the distribution of the gas that is accreted onto the galactic disc. When the gas is not rotating, it tends to accrete more in the central regions of the galaxy because of its negligible angular momentum. This creates a high concentration of gas in the centre that is able to cool down and generate a higher SFR, that strongly depends on the gas density.

\begin{figure}
	\includegraphics[width=\columnwidth]{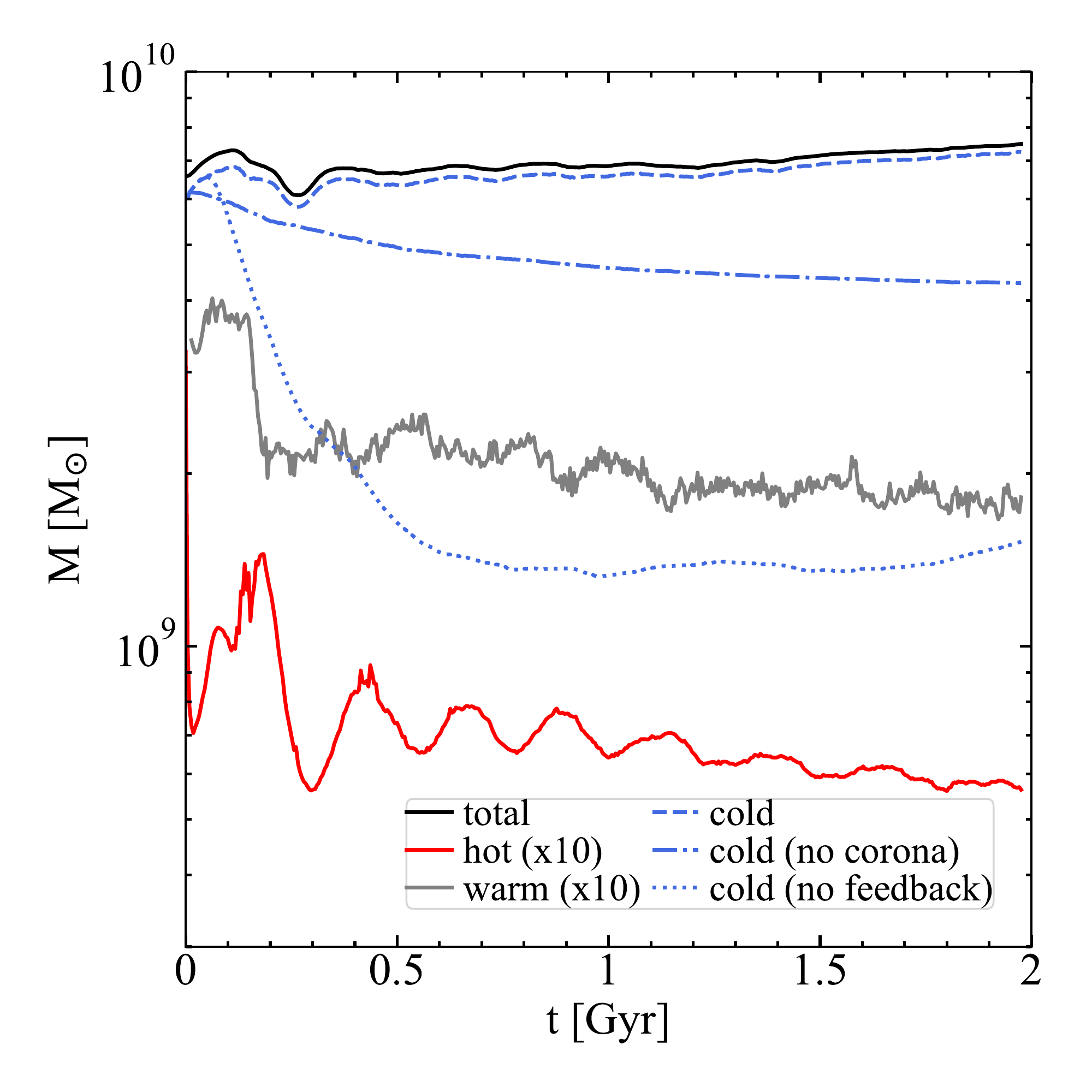}
	
	\caption{Mass of the gas in {\sc BM\_r} as a function of time at a height |z| $< 2$ kpc and a cylindrical radius R$<30$ kpc, divided in three ranges of temperature: cold ($T<10^4$ K, blue dashed line), warm ($10^4\le T \le 5\times10^5$ K, grey line) and hot ($T>5\times 10^5$ K, red line). The mass of warm and hot gas is multiplied by 10 for clarity. The black line represents the total mass of gas in {\sc BM\_r}. The blue dash-dotted line is the
mass of cold gas in {\sc nocor} and the blue dotted line the mass of cold gas in {\sc nofeed\_r}. The gas in the disc is almost totally cold, the warm and hot gas masses are lower by almost two orders of magnitude. In {\sc BM\_r} the mass of cold gas is approximately constant, while in {\sc nocor} it decreases over time. In {\sc nofeed\_r} the mass decreases at the beginning, reaching a constant value after 0.5 Gyr.}
    \label{mass_disc}
\end{figure}

In the {\sc nofeed\_r} simulation there is a large burst of star formation in the first 0.5 Gyr, reaching $50 \ M_{\odot} \ \text{yr}^{-1}$. In this time-span a gas mass of $\sim 10^{10}$ M$_{\odot}$ is converted into stars, depleting the disc of star-forming material. The following SFR trend is comparable to the one in the {\sc BM\_r} simulation, obtaining also in this case a SFR that stays more or less constant. For this reason, one might think that the stellar feedback is not particularly important for the evolution of the galaxy, but in the third row of Figure \ref{face1} we have seen the unrealistic disc structure in the {\sc nofeed\_r} simulation.  In absence of stellar feedback a few large cavities are generated, whereas high density filaments are absent and the gas is clustered in small clumps in the disc. Probably, the obtained SFR comes from the fact that in absence of stellar feedback we do not have negative feedback (which would tend to lead to a lower SFR) but neither positive feedback (which would tend to raise the SFR), given by the interaction between the galactic fountains and the corona. Despite the galactic fountains absence, the gas is still slowly accreted from the corona onto the disc. The balance between these two types of feedback combined with an accretion of gas from the corona thus gives rise to an approximately constant and similar SFR in the cases with and without feedback.

The {\sc nocor} simulation features a SFR that goes from $\approx 3 \ M_{\odot} \ \text{yr}^{-1}$ to $\approx 0.5 \ M_{\odot} \ \text{yr}^{-1}$ at 2 Gyr. We turn our attention on the differences given by the presence of the coronal gas. In {\sc BM\_nr}/{\sc BM\_r} the SFR is higher and is declining more slowly, making the galaxy able to maintain an almost constant/slightly decreasing SFR. This behaviour is caused by the accretion of the coronal gas onto the disc, that keeps high the density in the disc (maintaining in particular high density filaments, see Figure \ref{face1}), and, as a consequence, the SFR than in the no accretion scenario. This points out how the cold gas reservoir in the disc, that is used to form stars, is rapidly consumed if not replaced by an external source of gas, as expected from the depletion time (see Equation \ref{deptime}).

\begin{figure*}
	\includegraphics[width=\textwidth]{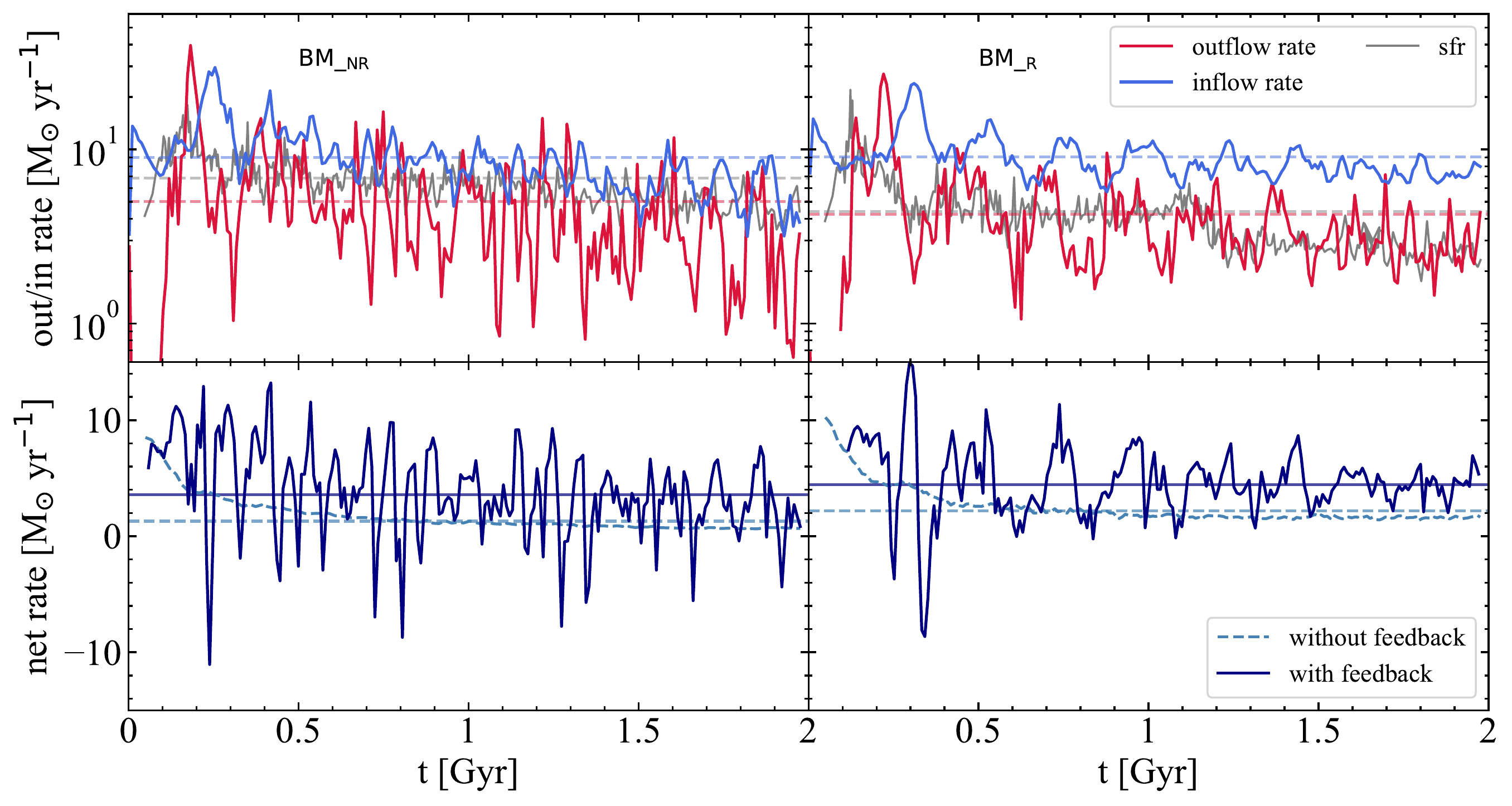}
    \caption{\textit{Top row:} Outflow (red line) and inflow (blue line) rates computed for the {\sc BM\_nr} (left-hand panel) and {\sc BM\_r} (right-hand panel) simulations, from gas inside slabs with a 300 pc width located at $\pm 2$ kpc from the plane of the disc. The grey dashed line represents the SFR. The three horizontal dashed lines represent the average value of the outflow rate (orange), inflow rate (light blue) and SFR (grey). The inflow and outflow rates highlight a strict connection between the formation of stars and the circulation of gas in the galaxy. \textit{Bottom row:} net inflow rate for the same two simulations (solid blue lines), considering also two cases without feedback processes (dashed light blue lines): {\sc nofeed\_nr} (left-hand panel) and {\sc nofeed\_r} (right-hand panel). The net inflow rates are computed as $\dot{M}_{\text{net}}=\dot{M}_{\text{in}}(t)-\dot{M}_{\text{out}}(t-t_{\text{cycle}})$, with $t_{\text{cycle}}=100$ Myr. The average net inflow rates are depicted with horizontal lines. We notice that more gas (about twice as much) is accreted if the stellar feedback is present in the simulation.}
    \label{outflow}
\end{figure*}

To further understand how this phenomenon occurs, in Figure \ref{mass_disc} we show the mass of gas in the disc (gas cells with $|z|<2$ kpc and $R<30$ kpc) in {\sc BM\_r} as a function of time. Gas has been divided in cold ($T<10^4$ K, blue dashed line), warm ($10^4\le T\le 5\times 10^5$ K, grey line) and hot phases ($T>5\times 10^5$ K, red line). The black solid line represents the total gas mass in the disc. As a comparison, the dash-dotted and dotted blue lines represent the mass of cold gas in {\sc nocor} and {\sc nofeed\_r} simulations, respectively. In general, inside the disc, the majority of the gas is cold, but there is also a small fraction ($\sim 10^8$ M$_{\odot}$) of warm and hot gas. The mass of cold gas starts from $\sim 6 \times 10^9$ M$_{\odot}$ and is $\sim 7 \times 10^9$ M$_{\odot}$ at the end of the simulation. Therefore, the mass of gas is not decreasing over time, but is rather increasing slightly, which would not be possible without the accretion of gas from the corona. In fact, in the {\sc nocor} case the mass of cold gas decreases over time down to $\sim 4 \times 10^9$ M$_{\odot}$ after 2 Gyr, since it is consumed to form stars without being replaced. This is in line with what we expected, with an average SFR of $1 \ M_{\odot} \ \text{yr}^{-1}$ a gas mass of $2 \times 10^9$ M$_{\odot}$ is consumed. In {\sc nofeed\_r} the mass of cold gas decreases rapidly in the first 0.5 Gyr, the negative feedback is absent and stars are formed at a rate of over $10 \ M_{\odot} \ \text{yr}^{-1}$, which causes the disc to consume $4.5 \times 10^9$ M$_{\odot}$ in the time probed by the simulation. After that, the mass remains approximately constant, although with $5.5\times10^9$ M$_{\odot}$ less than in the presence of feedback. Despite the much smaller mass of gas, the SFR has a trend similar to {\sc BM\_r} (Figure \ref{SFH}), the difference is in how the gas is distributed in the disc, in {\sc nofeed\_r} the gas can cluster in a few areas (in the centre or in other clumps, Figure \ref{face1}), increasing the efficiency of the gas-stars conversion process.
The ability of the galaxy, in {\sc BM\_nr}/{\sc BM\_r}, in maintaining a higher SFR longer than in {\sc nocor} can be attributed to the presence of the galactic corona that, accreting onto the disc, replenishes the disc with cold gas and slows down the emptying of the cold gas reservoir in the disc.

\begin{figure*}
	\includegraphics[width=\textwidth]{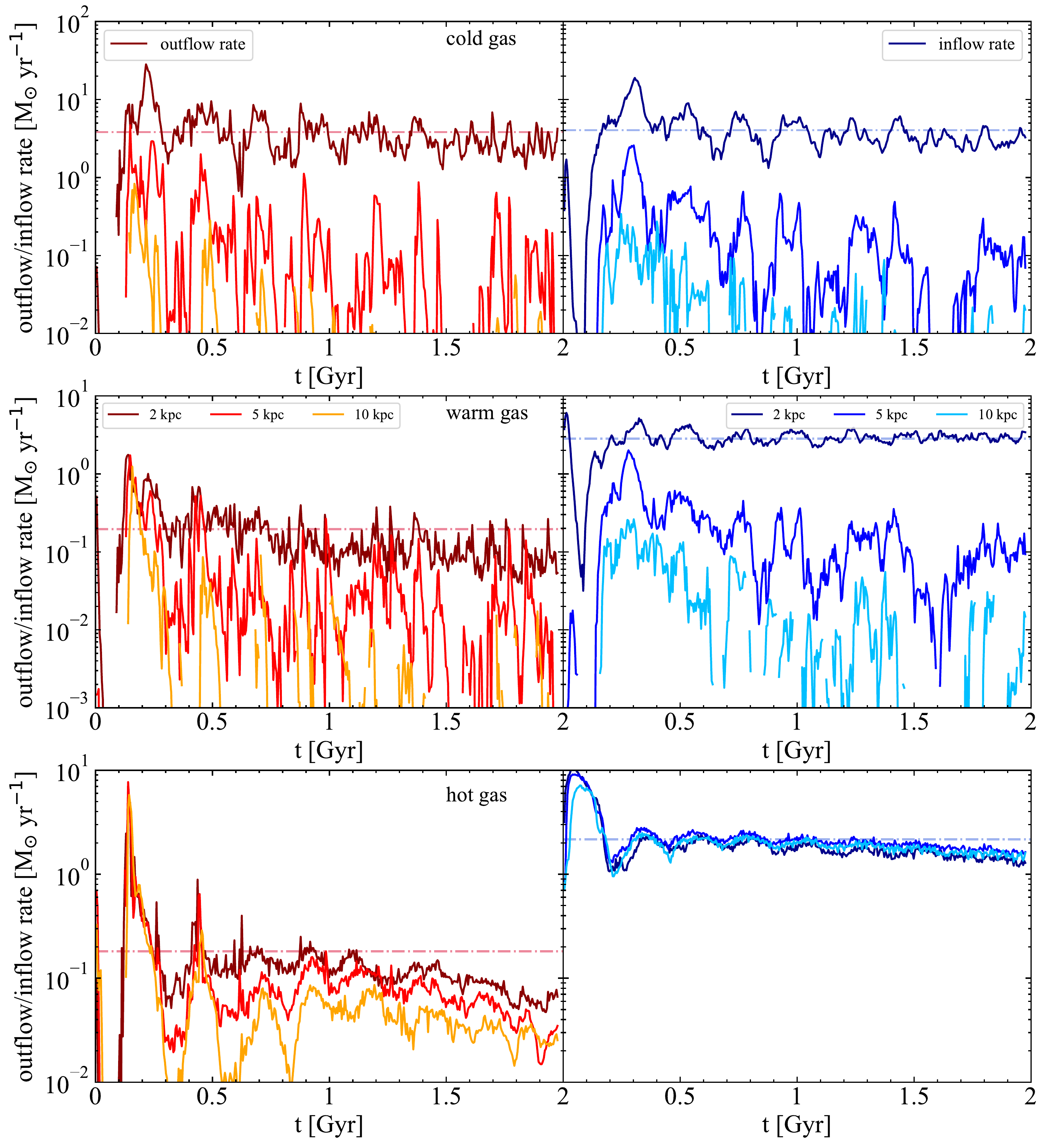}
    \caption{Outflow (red lines) and inflow (blue lines) rates for the {\sc BM\_r} simulation for three ranges of temperature at three different heights from the plane of the disc: 2 kpc, 5 kpc and 10 kpc (represented with three shades of red or blue, as described in the legend). The horizontal dashed lines represent the average value of the outflow/inflow rates at 2 kpc. \textit{Top panel}: cold gas ($T < 10^4$ K), the inflow and outflow rate values are almost identical, this phase seems to be generated for the majority by the galactic fountains ejected from the disc owing to stellar feedback. \textit{Middle panel}: warm gas ($10^4 \le T \le 5\times10^5$ K), the outflow rate is low ($\approx 0.2 \ M_{\odot} \ \text{yr}^{-1}$) with respect to the inflow rate. This gas is formed at the disc-corona interface, the same region where the galactic fountains are present, therefore we speculate that it might be created by the interaction between the corona and the cold gas. \textit{Bottom panel}: hot gas ($T > 5\times10^5$ K), the accretion of this gas derives from the cooling of the galactic corona and the values of the inflow rate at different heights from the disc is very similar.}
    \label{outtemp}
\end{figure*}

\begin{figure}
	\includegraphics[width=\columnwidth]{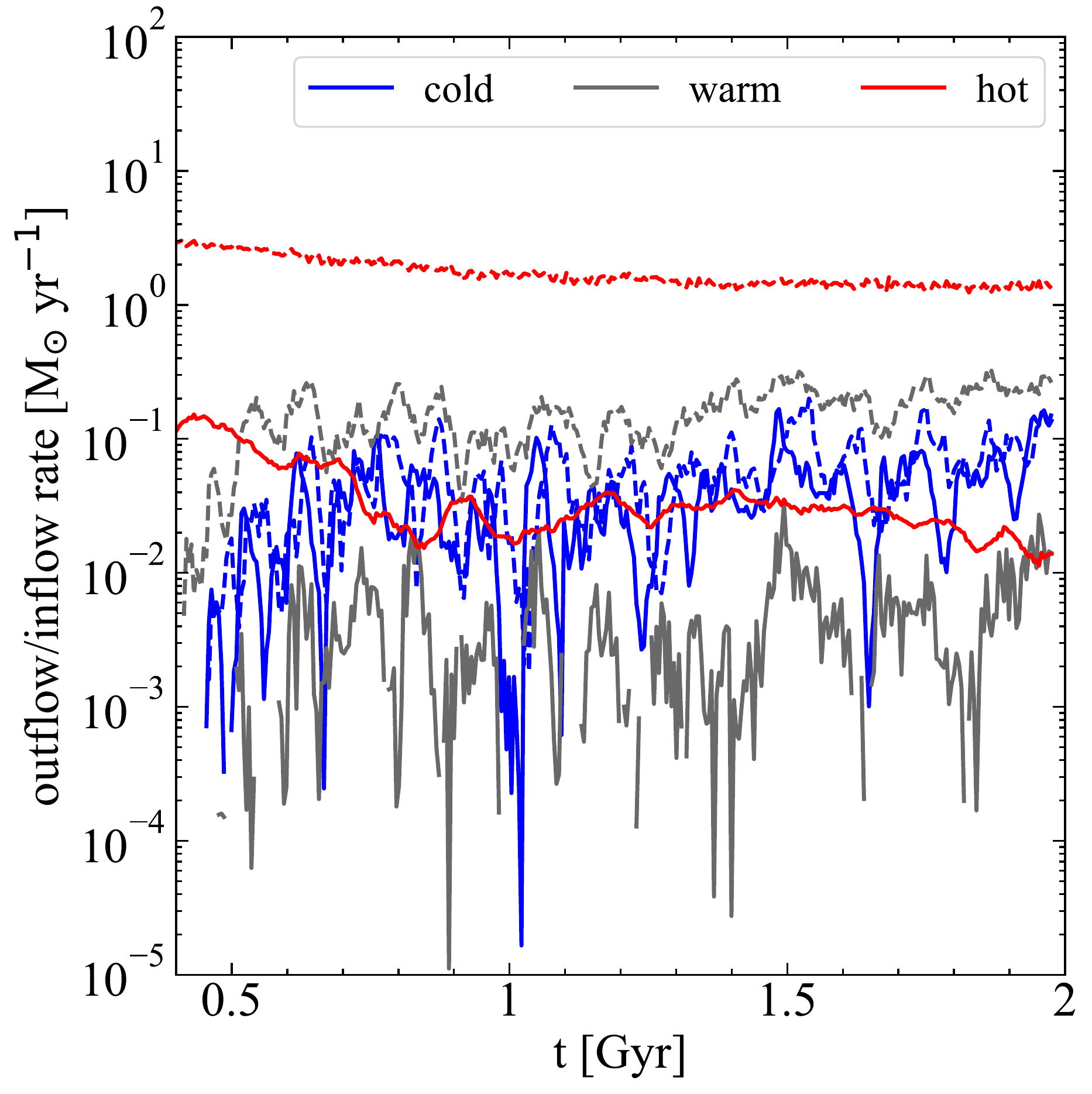}
	\caption{Outflow (solid lines) and inflow (dashed lines) rates of the simulation {\sc nofeed\_{r}}\ computed from gas inside slabs with a 300 pc width located at an height $|z| = 2$ kpc from the plane of the disc for the same three ranges of temperature of Figure \ref{outtemp}: cold (blue lines), warm (grey lines) and hot (red lines) gas. Only the inflow rate of hot gas has a non negligible value and there is almost no accretion of cold and warm gas. Furthermore, the outflow rates are also almost zero, due to the absence of stellar feedback no outflows are generated in the simulation.}
    \label{out_nofeed}
\end{figure}

\subsection{Galactic outflows and inflows}\label{outin}

To understand the relationship between star formation and the gas ejected/accreted from/onto the disc of the galaxy, we analyse the outflow and inflow rates of gas, schematically depicted in Figure \ref{galaxy_scheme}. We define the total inflow rate as
\begin{equation}
 \dot{M}_{\text{in}}=\dot{M}_{\text{in, fount}}+\dot{M}_{\text{in, int}}+\dot{M}_{\text{in, cor}}.
\end{equation}

We computed inflow and outflow rates as described in \citet{marinacci2019simulating}: we took two slabs of width  $\Delta z = 0.3$ kpc starting at a height from the plane of the disc of $\pm 2$ kpc (see the shaded area around the dashed horizontal line in Figure \ref{face1}) and sum up the contribution of all gas particles to the outflow/inflow rate
\begin{equation}
    \dot{M}_{\text{out/in, i}}=\frac{v_{z,i} m_i}{\Delta z},
\end{equation}
within the slab and with $R<30$ kpc. In the above equation $v_{z,i}$ is the gas particle vertical velocity and $m_i$ is the gas particle mass. The outflow rate is given by the gas particles that are moving away from the disc, i.e. particles with $v_zz>0$; while the condition $v_zz<0$ identifies the particles corresponding to an inflow. 

In the top panels of Figure \ref{outflow} we show the outflow (red line) and the inflow (blue line) rates at a height of 2 kpc over the disc plane for the {\sc BM\_nr} (left-hand panel) and {\sc BM\_r} (right-hand panel) simulations. The grey line represents the corresponding SFR. The three horizontal dashed lines represent the average values of the inflow (light blue) and outflow (orange) rates and the average SFR (grey).

In {\sc BM\_nr} the inflow rate ($\langle\dot{M}\rangle_{\text{in}} \approx 9 \ M_{\odot} \ \text{yr}^{-1}$) is, on average, higher than the outflow rate ($\langle\dot{M}\rangle_{\text{out}}\approx 5 \ M_{\odot} \ \text{yr}^{-1}$). The rates in {\sc BM\_r} are very similar, the average inflow and outflow rates are $\langle\dot{M}\rangle_{\text{in}}\approx 9 \ M_{\odot} \ \text{yr}^{-1}$ and $\langle\dot{M}\rangle_{\text{out}}\approx 4 \ M_{\odot} \ \text{yr}^{-1}$ respectively.
It is worth noticing that in both simulations most of the peaks in the inflow rate correspond to depressions in the outflow rate and vice versa. This indicates the presence of a gas circulation at the interface between the disc and the halo (occupied by the corona) of the galaxy: the stellar feedback produces galactic fountains (outflows) that pass through the corona and then fall back onto the disc (inflows). We expect this phenomenon to drag also some coronal gas that could refill the disc and sustain the star formation.

Looking at the SFR, the first noticeable thing is that in both {\sc BM\_nr} and {\sc BM\_r} the average SFR ($\approx 6.8 \ M_{\odot} \ \text{yr}^{-1}$ in {\sc BM\_nr} and $\approx 4.4 \ M_{\odot} \ \text{yr}^{-1}$ in {\sc BM\_r}) is below the average inflow rate, implying that the supply of gas from the corona has a non negligible influence on the total gas budget of the galaxy, making it an important reservoir for the star formation. Taking these numbers at face value, we can infer that star formation can be sustained entirely by the accretion of the coronal gas. Indeed, the inflowing gas is mostly caused by the accretion due to coronal gas. Computing the average net inflow rate as $\langle \dot{M}_{\text{net}}\rangle =\langle\dot{M}\rangle_{\text{in}}-\langle\dot{M}\rangle_{\text{out}}$, we obtain $\approx 4 \ \text{M}_{\odot} \ \text{yr}^{-1}$ in {\sc BM\_nr} and $\approx 5 \ \text{M}_{\odot} \ \text{yr}^{-1}$ in {\sc BM\_r}, it follows that only a fraction of the inflows is accounted for by the gas that is ejected from the disc and that returns back on it ($\dot{M}_{\text{in, fount}}$), the rest of the gas must be accreted from the corona ($\dot{M}_{\text{in, int}}+\dot{M}_{\text{in, cor}}$). 

On average, the inflow rate in {\sc BM\_nr} and {\sc BM\_r} is almost identical. Therefore, the rotation of the coronal gas is not decreasing the rate at which the gas is falling onto the disc, but rather the distribution of this accreted gas appears to be different. The lower SFR in {\sc BM\_r} (Figure \ref{SFH}) tells us that the accreted gas is less efficient in forming new stars with respect to {\sc BM\_nr}. One hypothesis is that the gas is more easily accreted in the outermost areas of the disc rather than in the centre, where the density of the gas is higher and therefore stars can be formed more easily.

Furthermore, we note that the SFR trend is linked to the outflow/inflow rates. Intervals of above average SFR are often followed by local outflow rate peaks and reciprocally the SFR peaks follow the inflow rate peaks. The presence of this correlation between the SFR and outflow/inflow rate trends highlights the role of the star formation in the circulation of gas in the galaxy. The gas accreted from the corona can efficiently cool and form stars in the disc of the galaxy, which subsequently expel a fraction of gas from the disc. This underlines a strict connection between these phenomena, showing how the star formation is intimately linked to the accretion/ejection of gas from/into the external environment.

\subsection{Positive feedback scenario}\label{pos_feed}

One fundamental phenomena that has to be understood is whether or not the stellar feedback might act as a positive feedback for star formation. This has been already studied in the literature \citep[e.g.][]{marinacci2010, hobbs2020}, and it is speculated that the galactic fountains originated from the stellar feedback can enhance the accretion of gas from the galactic corona, owing to the mixing of the gas ejected from the disc with the hot gas that surrounds it. This can happen as a result of turbulence or of the generation of a Kelvin-Helmoltz instability in the wake of the galactic fountains. This newly generated mixture of gas has a lower cooling time with respect to the original coronal gas, ranging from a few hundred Myr to $\lesssim 1$ Myr \citep{marinacci2010} due to the lower temperature (between $10^4$ and $10^6$ K) and higher metallicity (between 0.1 and 1 $Z_{\odot}$), allowing a more rapid accretion. 

To better investigate this phenomenon, to which we will refer as positive feedback scenario, we analyse the set of simulations without stellar feedback {\sc{nofeed\_nr}}/{\sc{nofeed\_r}} (see Table \ref{model_var}), these have the same ICs of {\sc{BM\_nr}}/{\sc{BM\_r}}, but the stellar feedback processes are disabled. In this way, there is no injection of energy and momentum in the ISM, therefore no outflow is produced and there is almost no interaction between the disc and the coronal gas in the extra-planar region. We want to compare {\sc{BM\_nr}}/{\sc{BM\_r}} to {\sc{nofeed\_nr}}/{\sc{nofeed\_r}} in order to see if there are eventually differences in the accretion of gas. In doing so, we have to consider only the gas accreted from the galactic corona. We want to understand how much gas is accreted from this component without the galactic fountains (that are not present without stellar feedback) contribution (see Figure \ref{galaxy_scheme}). In fact, the galactic fountain gas circulates outside and inside the galaxy, but does not change the quantity of gas stored in the disc, that can increase consequently to inflows of gas from the corona. Therefore, we need to subtract from the total inflow rate $\dot{M}_{\text{in}}$ the inflow rate coming from the galactic fountains $\dot{M}_{\text{in, fount}}$, that is $\sim \dot{M}_{\text{out}}$. For this reason, the net inflow rate has been computed as
\begin{equation}\label{tcycle}
    \dot{M}_{\text{net}}=\dot{M}_{\text{in}}(t)-\dot{M}_{\text{out}}(t-t_{\text{cycle}})
\end{equation}
at each time $t$. As depicted in Figure \ref{galaxy_scheme}, a large fraction of the inflowing gas is coming from the galactic fountains, the gas is ejected and then, after a typical time $t_{\text{cycle}}\sim 100$ Myr \citep{fraternali2014}, which is computed considering an average initial velocity of $70-80$ km s$^{-1}$ \citep{marasco2012}, falls back onto the disc. This time is in line with what we found in this work, measuring the average time span between two consecutive outflow/inflow rate peaks. Choosing a different $t_{\text{cycle}}$ in Equation (\ref{tcycle}) does not change significantly the results of following analysis.

In the lower panels of Figure \ref{outflow} we show the net inflow rates for the couple of simulations {\sc{BM\_nr}}/{\sc{nofeed\_nr}} (left-hand panel) and {\sc{BM\_r}}/{\sc{nofeed\_{r}}} (right-hand panel) computed as in Equation (\ref{tcycle}).
The horizontal lines represent the average value of the net inflow rate with (solid blue line) and without (dashed light blue line) stellar feedback. We can see that the mean net inflow rate without feedback is lower than the one with feedback in both cases. In the {\sc{nofeed\_nr}} simulation the rate is $\dot{M}_{\text{net}}\approx$\,1.3 \,{\rm M}$_{\odot}$ yr$^{-1}$. Without the stellar feedback $\dot{M}_{\text{net}}$ is approximately half of the net inflow rate in {\sc{BM\_nr}} ($\approx 3.6\, {\rm M}_{\odot}$ yr$^{-1}$). Essentially the same applies for the {\sc BM\_r}/{\sc{nofeed\_r}} simulations, in which the net rates are $\approx \, 2.2 \,{\rm M}_{\odot}$ yr$^{-1}$ and $\approx 4.5\, {\rm M}_{\odot}$ yr$^{-1}$ respectively. The rates found without feedback are in line with a rough estimate of the accretion rate computed as the mass contained in the corona cooling radius (i.e. the radius at which the corona cooling time equals the simulation time span), which is about $15-20$ kpc. Therefore, it is the mass of gas that we expect to cool in 2 Gyr in a pure radiative cooling scenario. The efficiency of gas accretion from the corona to the disc is higher if stellar feedback is present. Therefore, stellar feedback is increasing the rate at which the gas is falling onto the disc. This is due to the mixing between the two gas phases (the cold ISM and the hot corona) induced by the stellar feedback itself.

\subsection{Gas phases at the disc-corona interface}\label{outt}

The gas contained in the outflows and inflows is a complex medium composed of different phases. Understanding their evolution may allow us to better comprehend how the circulation of gas around the galaxy and the positive feedback scenario work. In Figure \ref{outtemp}, where we consider only our fiducial simulation {\sc BM\_r}, we show the outflow (red shades) and inflow (blue shades) rates in three ranges of temperatures, cold ($T < 10^4$ K, first panel), warm ($10^4 \le T \le 5\times10^5$ K, second panel) and hot ($T > 5\times10^5$ K, third panel) and at three heights from the plane of the disc (2 kpc, 5 kpc and 10 kpc, see the dashed horizontal lines in Figure \ref{face1}), highlighted by the three shades of colour. These rates are computed as described in Section \ref{outin}.

In the top panel of Figure \ref{outtemp}, we show the cold phase. The cold ($T < 10^4$ K) outflow and inflow rates have a very similar trend, but shifted on average by $t_{\text{cycle}}\sim 100$ Myr, which is the average time that a galactic fountain takes to return to the disc after the ejection, as already mentioned in Section \ref{outin}. The average value of the outflow and inflow rates is also similar ($\approx 4$ M$_{\odot}$ yr$^{-1}$) and this gives us information on the origin of this phase of gas. We speculate that it represents the gas from the galactic fountains, that are mainly made by the cold gas of the disc: the gas that is ejected from the plane of the disc is re-accreted after a fountain cycle, this explains the periodic shifted trend of the ejected and accreted gas. This is corroborated by the fact that both outflow and inflow rates at 5 and 10 kpc are substantially lower. In fact, the majority of the fountains stays in the first kpc from the disc, as it can be seen from the gas projections (see Figures \ref{face1}, \ref{edgevel}, \ref{temp_proj}) only the strongest fountains can reach heights $|z|>10$ kpc.

The middle panel shows the warm phase ($10^4 \le T \le 5\times10^5$ K). This is particularly important for the positive feedback scenario as we can argue that the warm gas is generated from the mixing between the cold galactic fountains and the hot corona. In fact, in this case there is a substantial difference between the gas that is accreted and the gas that is ejected. The inflow rate is much higher ($\approx 3$ M$_{\odot}$ yr$^{-1}$) with respect to the outflow rate, that is almost negligible ($\approx$0.2 M$_{\odot}$ yr$^{-1}$), only a small fraction of the galactic fountains is composed by warm gas. Therefore, this gas phase is not lifted from the disc, but is rather generated at the disc-corona interface. The inflow rate of warm gas is lower at 5 and 10 kpc (below 0.2 M$_{\odot}$ yr$^{-1}$), similarly to the cold phase. Hence, it seems to be generated from the same region where the galactic fountains are present. This warm gas seems to be created for the most part in the first 2-3 kpc over the disc, and is then accreted onto the disc giving a non-zero inflow rate. This indicates how the accretion of warm gas is not generated by a cooling flow but is rather created due to the positive action of stellar feedback.

The bottom panel shows the hot gas ($T > 5\times10^5$ K). Only a small part of the ejected gas is made by this phase (coming from the gas shock-heated from the SNe), giving a outflow rate of $\approx 0.2$ M$_{\odot}$ yr$^{-1}$. The inflow rate is stable at a value of $\approx 2$ M$_{\odot}$ yr$^{-1}$ at the three different heights. Therefore, the accretion has the same intensity at 2, 5 and 10 kpc over the plane of the disc. This gas is probably part of the galactic corona that is accreted onto the galaxy and it moves towards the central regions of the galaxy replacing the gas accreted onto the disc. 

 An interesting trend that can be noticed is that the gas with higher entropy ($s\propto T/\rho^{2/3}$) reaches higher distances from the disc than the cold gas. Since a gas distribution is convectively stable if entropy increases with radius, hotter and less dense gas can circulate towards higher distances with respect to cold gas. This behaviour is apparent if we look at the outflowing gas in Figure \ref{outtemp}: only a small fraction of the cold ejected gas ($\sim 0.5 \%$) arrives at 10 kpc. This fraction increases with the temperature and is $\sim 10 \%$ for the warm gas and $\sim 60 \%$ for the hot gas.

 One may argue that the intermediate warm phase of the accreted gas could come from the radiative cooling of the corona without the presence of cold galactic fountain gas that may act as a catalyst, and that this can happen near the disc where the density and the cooling function of the corona are higher. To rule out this possibility we have computed the same outflow/inflow rates at 2 kpc, shown in Figure \ref{out_nofeed}, in the {\sc{nofeed\_r}} simulation, where the cooling of the corona is present, but the stellar feedback is turned off. In this case the outflow rates (solid lines) are negligible because of the feedback absence. The only important accretion of gas (dashed lines) is in the hot phase (red colour) at rates of $2-3 \ M_{\odot}$ yr$^{-1}$, that are very similar to the {\sc BM\_r} simulation. The accretion of cold (blue colour) gas is below $0.1 \ M_{\odot}$ yr$^{-1}$ because the galaxy is lacking the galactic fountains. A similar rate is present for the warm phase (grey colour), the absence of ejections of cold gas from the disc means that the warm gas cannot be formed. 
 
 Hence, we found that the warm gas found in our simulations is generated as a consequence of a gas circulation at the disc-corona interface, mediated by the galactic fountains ejected from the disc. This warm gas then can fall towards the galactic disc, fueling the galaxy with a non-negligible ($\approx 3$ M$_{\odot}$ yr$^{-1}$) amount of gas, that is necessary to sustain the SFR found in {\sc BM\_r} ($\approx 4$ M$_{\odot}$ yr$^{-1}$).

\section{Comparison to previous works}\label{caveats}

In the last years different numerical works have been concerned with the study of the galactic corona, or with the circumgalactic medium in general, analysing its properties \citep[e.g.][]{lochhaas2020}, the relation with outflows \citep[e.g.][]{pandya2021} and numerical cooling flow solutions \citep[e.g.][]{stern2019,stern2020}. In particular, \citet{stern2020} hypothesizes that under a critical accretion rate $\dot{M}_{\text{crit}}$ of $\approx 10 \ {\rm M_{\odot}} \ \text{yr}^{-1}$ for a Milky Way-like galaxy, the gas is accreted into the disc nearly at the virial temperature ($\sim 10^6$ K). This is consistent with our results in absence of feedback ({\sc nofeed} simulations), in fact, in this case in our simulations the accretion rate from the corona is lower than $\dot{M}_{\text{crit}}$ and therefore it happens in the hot mode (see red dashed line in Figure \ref{out_nofeed}), with a gaseous flow that stays hot until it reaches the interface region with the disc. In the simulations including feedback this hot gas can eventually mix with the gas ejected by stellar feedback and its accretion rate is enhanced.

Comparatively few works have focused on the interaction between the galactic fountains and the corona, and on the resulting positive feedback \citep[e.g.][]{marinacci2010,marinacci2011,armillotta2016,hobbs2020,gronke2022}. We focus, in particular, on \citet{marinacci2010} and \citet{hobbs2020} whose aim is more similar to our work. Both papers found evidence of positive feedback coming from this galactic fountain-corona interaction.

\citet{marinacci2010} developed 2D simulations, with the Eulerian code ECHO, of a galactic fountain cloud with $T=10^4$ K moving through the galactic corona ($T=2\times 10^6$ K), with an initial velocity of $75$ km s$^{-1}$. They found that for reasonable density ($n\approx2\times10^{-3}$ cm$^{-3}$) and metallicity ($Z\approx 0.1$ Z$_{\odot}$) values of the corona, some coronal gas can condensate in the wake of the cloud, leading to a mass of cold gas that increases over time as the fountain passes through the hot gas. By extrapolating the result for a single fountain gas cloud to the whole galactic disc, they obtained a global accretion rate of $\approx 0.5$ M$_{\odot}$ yr$^{-1}$ for the Milky Way.

\citet{hobbs2020} investigated the positive feedback process simulating, with the {\sc Gizmo} code, a portion of a star-forming disc in a small box of  $2\times2\times \pm 50$ kpc; therefore the resolution ($23$ M$_{\odot}$ for gas particles) is higher than in our case. The galactic corona has a constant number density of $6\times10^{-4}$ cm$^{-3}$ and a constant temperature of $\approx 10^6$ K. They also take into account gas turbulence in the gaseous disc and use a different stellar feedback implementation (see \citealt{kim2014}), in which SN feedback has a purely thermal component. Despite the differences in the ICs and in the models, their simulations suggest the presence of a positive stellar feedback mechanism as well. Furthermore, also in their case the mass of cold gas in the disc remains approximately constant as a consequence of coronal accretion. \citet{hobbs2020} developed an analytical model to examine the dependence of the accretion rate on the SFR. They found that negative feedback starts dominating over positive feedback when the SFR (extrapolated to the whole galactic disc) overcomes a value of $\approx 0.5 \ {\rm M_{\odot}} \ \text{yr}^{-1}$. Above this value, an increase in the SFR will produce a decrease in the accretion rate. 
Although we would need more galaxy configurations to robustly investigate this aspect, it seems that our simulations behave differently as there is not an obvious trend between an increase/decrease of the SFR and a corresponding decrease/increase in the accretion rate (see Figure \ref{outflow}). Our simulations appear to suggest that for a Milky Way-like galaxy positive stellar feedback continues to be important even at SFR above $\approx 0.5 \ {\rm M_{\odot}} \ \text{yr}^{-1}$.

\section{Summary and Conclusions}\label{conclusions}

In this work we studied the evolution of a series of multicomponent $N$-body hydrodynamical models representative of a Milky Way-like galaxy embedded in a hot gaseous atmosphere, the so-called corona, using state-of-the-art numerical simulations. In particular, we focused on the interaction between the disc and the corona caused by the gas exchanges between these two components mediated by stellar feedback. We studied several aspects of this interaction, such as the balancing between the star formation and the outflows/inflows of gas and the mixing between the disc and the corona. We analysed these aspects employing the moving-mesh code \arepo\ and the explicit stellar feedback and ISM medium model \smuggle\ to carry out our simulations. We have successfully built a set of ICs for the hydrodynamical $N$-body simulations analysed in this work. These configurations have then been used to perform `full-physics' simulations with the \smuggle\ model. The main results of these calculations can be summarized as follows.
\begin{itemize}
    \item We found that the galactic corona is the main contributor to the material sustaining star formation in the galaxy. The star formation takes place from the cold gas in the disc, whose mass is maintained at a nearly constant level by gas accretion from the corona; without this process the galaxy would have a lower SFR and would progressively deplete its gas content and lower its star formation rate in $\sim 1-2$ Gyr.
    
    \item We studied the outflow/inflow rates from/onto the galactic disc. We have observed the formation of gaseous outflows due to the combined effects of the different stellar feedback processes implemented in the \smuggle\ model. The analysed outflow rates are generated by the galactic fountains, which are composed mostly by cold metal-rich gas ($T<10^4$ K), but possess also a fraction of warm and hot gas ($T\ge 10^4$ K). The alternation of the peaks in the outflow and inflow rates, for which each outflow-dominant phase is followed by an inflow-dominant phase, suggests the presence of a circulation of the gas between the galactic disc and the corona, in fact almost $50\%$ of the inflow rate derives from the galactic fountains. The remaining inflow rate fraction is gas accreted onto the disc from the corona. We found a connection between the gas inflowing into the galaxy and star formation: the average SFR is below the average inflow rate, making the accretion of gas sufficient to entirely sustain star formation.
    
    \item We have analysed the \textit{positive feedback scenario}: the accretion of gas might potentially be helped by the interaction between the galactic fountains and the coronal gas, favoring its cooling. In doing so, we made a comparison between simulations with and without stellar feedback. After an initial starburst phase in the no feedback run, caused by an overcooling of the gas in absence of feedback, the SFR stabilizes to a level similar to the simulations that include stellar feedback. We explained that as a consequence of the absence of both positive and negative feedback, those phenomena compensate each other, raising and reducing the SFR. We found that stellar feedback helps in increasing the accretion rate of the coronal gas, which is two times larger than in the no feedback case. This increase of accretion is due to the generation of a warm gas phase (with an inflow rate of $\approx 3 \ M_{\odot} \ \text{yr}^{-1}$) at the disc-corona interface thanks to the action of positive stellar feedback. This warm gas then falls onto the disc, cools down and eventually fuels new star formation.
\end{itemize}

Our simulations currently do not include magnetic fields, that could lead to a different dynamics of gas and have an effect on the star formation efficiency and feedback, and accurate cosmic rays dynamics, that might have an effect on the global outflow rate and regulate the thermal and ionization state of the gas\footnote{However, we remind that a crude estimate of the contribution of cosmic rays to the heating of the gas is already included in \textit{SMUGGLE}.}. Including them in future work can give us further insights on behavior of the interstellar and circumgalactic media in star-forming galaxies. The type of simulations analysed in this work represents a bridge between idealized small-scale simulations and cosmological simulations, allowing to obtain a high/intermediate resolution while still considering a realistic galactic setup. Nevertheless, the model galaxy evolves in isolation and it is not inserted in a full cosmological context. Therefore, it would be desirable to extend this analysis to simulations including the cosmological context self-consistently. In this way, it will be possible to follow the evolution of star-forming galaxies like the Milky Way with a higher degree of physical fidelity, drawing a coherent picture of the formation and evolution of such objects. This will enable us to make an important step forward towards a more physically faithful modelling of the evolution of star-forming galaxies.

\section*{Acknowledgements}

We acknowledge the use of computational resources from the parallel computing cluster of the Open Physics Hub (\url{https://site.unibo.it/openphysicshub/en}) at the Physics and Astronomy Department in Bologna. RP acknowledges support from PRIN INAF 1.05.01.85.01. FM wishes to thank Greg Bryan for stimulating discussions about this work. LVS is grateful for financial support from the NSF-CAREER-1945310 and NASA ATP-80NSSC20K0566 grants. MV acknowledges support through NASA ATP 19-ATP19-0019, 19-ATP19-0020, 19-ATP19-0167, and NSF grants AST-1814053, AST-1814259, AST-1909831, AST-2007355 and AST-2107724. PT acknowledges support from NSF grant AST-2008490 and NASA ATP Grant 80NSSC22K0716.

\section*{Data Availability}

The data underlying this article will be shared on reasonable request to the corresponding author.



\bibliographystyle{mnras}
\bibliography{biblio} 

\bsp	
\label{lastpage}
\end{document}